\documentclass[twoside]{article}

\usepackage{PRIMEarxiv}

\usepackage[utf8]{inputenc} % allow utf-8 input
\usepackage[T1]{fontenc}    % use 8-bit T1 fonts
\usepackage{hyperref}       % hyperlinks
\usepackage{natbib}

\usepackage{url}            % simple URL typesetting
\usepackage{booktabs}       % professional-quality tables
\usepackage{amsmath,amssymb,amsfonts}%
\usepackage{nicefrac}       % compact symbols for 1/2, etc.
\usepackage{microtype}      % microtypography
\usepackage{lipsum}
\usepackage{fancyhdr}       % header
\usepackage{graphicx}       % graphics
\usepackage[title]{appendix}%
\usepackage{floatrow}
\usepackage{gensymb}
\usepackage{multirow}
\usepackage{multicol}

\graphicspath{{media/}}     % organize your images and other figures under media/ folder

\title{Active learning of data-assimilation closures using Graph Neural Networks}

\author{
  Michele Quattromini \\
  Dipartimento di Meccanica, Matematica e Management \\
  Politecnico di Bari \\
  Bari\\
  LISN-CNRS\\
  Université Paris-Saclay\\
  Orsay\\
  \texttt{michele.quattromini@poliba.it} \\
  \And
  Michele Alessandro Bucci \\
  Safran Tech, Digital Sciences \& Technologies \\
  Magny-Les-Hameaux\\
  \texttt{michele-alessandro.bucci@safrangroup.com} \\
  \And
  Stefania Cherubini \\
  Dipartimento di Meccanica, Matematica e Management \\
  Politecnico di Bari \\
  Bari \\
  \texttt{stefania.cherubini@poliba.it} \\
  \And
  Onofrio Semeraro \\
  LISN-CNRS\\
  Université Paris-Saclay\\
  Orsay\\
  \texttt{onofrio.semeraro@universite-paris-saclay.fr} \\
}

\newcommand{\vdir}{\mathbf{u}}
\newcommand{\pdir}{p}
\newcommand{\ff}{\mathbf{f }}

\newcommand{\mean}[1]{\overline{#1}}
\newcommand{\fluct}[1]{#1'}
\newcommand{\invRe}{\dfrac{1}{Re}}

\newfloatcommand{capbtabbox}{table}[][\FBwidth]

\begin{document}
\maketitle

% Abstract
\begin{abstract}
    
The spread of machine learning techniques coupled with the availability of high-quality experimental and numerical data has significantly advanced numerous applications in fluid mechanics. Notable among these are the development of data assimilation and closure models for unsteady and turbulent flows employing neural networks (NN). Despite their widespread use, these methods often suffer from overfitting and typically require extensive datasets, particularly when not incorporating physical constraints. This becomes compelling in the context of numerical simulations, where, given the high computational costs, it is crucial to establish learning procedures that are effective even with a limited dataset.
Here, we tackle those limitations by developing NN models capable of generalizing over unseen data in low-data limit by: i) incorporating invariances into the NN model using a Graph Neural Networks (GNNs) architecture; and ii) devising an adaptive strategy for the selection of the data utilized in the learning process. GNNs are particularly well-suited for numerical simulations involving unstructured domain discretization and we demonstrate their use by interfacing them with a Finite Elements (FEM) solver for the supervised learning of Reynolds-averaged Navier--Stokes equations. We consider as a test-case the data-assimilation of meanflows past generic bluff bodies, at different Reynolds numbers $50 \leq Re \leq 150$, characterized by an unsteady dynamics. We show that the GNN models successfully predict the closure term; remarkably, these performances are achieved using a very limited dataset selected through an active learning process ensuring the generalization properties of the RANS closure term. The results suggest that GNN models trained through active learning procedures are a valid alternative to less flexible techniques such as convolutional NN.
\end{abstract}

%% Graphical abstract
%\begin{graphicalabstract}
%\includegraphics[width=1\textwidth]{images/graphical_abstract.png}
%\end{graphicalabstract}

%% Research highlights
%\begin{highlights}
%%
%    \item \textbf{Interface FEM-GNN}: This study introduces a novel approach that combines Graph Neural Networks (GNNs) with Finite Element Method (FEM) solvers. The goal is to tackle the closure problem in Reynolds-averaged Navier-Stokes (RANS) equations. The GNN architecture is designed to exploit distance information between nodes to enhance its ability to model the behavior of RANS equations.
    %
%    \item \textbf{Generalization capabilities with limited amount of data}: The research investigates the GNN model’s ability to generalize to unseen data. Notably, the GNN demonstrates reasonable prediction accuracy even with limited datasets, , in contrast with the typical data-hungry nature of traditional Neural Network models.
    %
%    \item \textbf{Data Efficiency and Active Learning Strategy}: The study further assesses the impact of both quantity and quality of training data on the GNN model’s generalization capabilities. An adaptive strategy based on the similarity of the gradients is introduced, guiding the selection of training data during the learning process to optimize the GNN model performance.
    %
%\end{highlights}

% keywords can be removed

\section{Introduction}
\label{sec:introduction} 
Machine learning (ML) applications are spreading across the scientific community in the most diverse fields, leading to the development of novel techniques and algorithms encompassed within the emerging field of {scientific machine learning}. This breakthrough is fostered by the massive augmentation of computer-aided activities, the constant increase of high performance computational power, the recent developments in the field of deep learning and the large availability of data \citep{goodfellow2016deep}. In a very schematic way, for a regression problem, ML techniques enable to identify mappings between observables of a system (inputs) and quantities of interest (outputs) we aim to predict by leveraging data; when these analysed data are governed by deterministic or statistical laws, in principle, these mappings correspond to approximating models. Alternative applications can be found in classification problems, clustering, control in a large variety of scientific fields, including fluid mechanics where the number of contributions combining ML and standard techniques of analysis has been constantly growing in the last years \citep{raissi2020hidden, brunton2020special, brunton2020machine, garnier2021review, vinuesa2022enhancing, yang2024enhancing}.

These applications found a natural ground into modelling problems, ranging from the closure of Reynolds-Average Navier-Stokes equations for turbulent flows, to wall model Large Eddy Simulations \citep{lapeyre2019training, zhou2021wall, bae2022scientific} or optimization problems combining adjoint-based techniques and machine learning \citep{volpiani2021machine, patel2024turbulence}. Among the numerous authors that addressed this problem, \cite{LingTempleton2015} applied classification methods for identifying regions of uncertainties where the closure term of the Reynolds-averaged Navier--Stokes (RANS) might fail; \cite{strofer2021end} combined data assimilation with neural networks (NN) modelling of the Reynolds stress using limited observation. Other approaches leverage baseline models such as the the Spalart-Allmaras closure \citep{singh2016using}, physics-informed NN (PINNs) \citep{eivazi2022physics}, regression methods \citep{schmelzer2019machine}, decision trees \citep{duraisamy2019turbulence}, ensemble methods \citep{mcconkey2022generalizability} or genetic programming \citep{weatheritt2016novel, zhao2020rans}. For a broader overview, we refer to \cite{duraisamy2019turbulence} and \cite{beck2021perspective}, where the different levels of approximation are discussed together with a critical take on the limitations of the approach.
In the specific case of an eddy viscosity closure model, the recent studies by \cite{ling2016reynolds} demonstrated the effectiveness of Tensor-Basis Neural Networks (TBNNs) in learning a General Eddy Viscosity model type \citep{pope1975more}. TBNNs capitalize on the tensor decomposition approach proposed by Pope to account for invariances and streamline the number of parameters to be learnt. The inductive bias introduced by this modelling approach restricts TBNN application to nearly homogeneous flows with high Reynolds numbers, where local effects predominate \citep{cai2024revisiting}. On the other hand, at low Reynolds numbers, we may consider as alternative approach a data assimilation one, where the closure model corresponds to the control parameter of an adjoint-based loop \citep{foures}; without explicitly introducing a tensor structure or Boussinesq approximations, this method is well suited for non-homogeneous flows at lower Reynolds numbers. Here, we take inspiration from these applications and mainly focus on unsteady flows developing past bluff bodies at low Reynolds numbers $50 \leq Re \leq 150$; we consider RANS as baseline, although alternative choices can be also adopted, such as Euler equations or linearized Navier Stokes equations in resolvent form, where the parameter to be tuned is the dissipation term \cite{morra2019relevance, pickering2021optimal,von2023self}. With respect of standard approaches, we focus here on a supervised learning strategy where the closure model is identified by inference from available data. In principle, we could identify universal closure models directly from data, having at disposal an infinite amount of them. In practice, in real cases, we may deal with a limited amount of data or few localized measurements; these limitations can impact on the use of methods such as neural networks (NN), where the expressivity and generic structure make them suitable for a large class of models, but prone to generalisation problems. In this sense, ML models risk to be representative solely of the datasets included in the training process; thus, it becomes compelling given the available data to circumvent these problems by inputting well selected data during the training or providing prior knowledge through modelling \citep{bucci2021leveraging, shukla2022scalable}. 

\smallskip
With this in mind, we will test if it is possible to identify generic closure models from data defined on unstructured meshes while only relying on a small amount of data chosen on principled criteria. In order to answer these questions, a first ingredient is the introduction of graph neural networks (GNN) \citep{hamilton2020graph}; this architecture is characterized by complex multi-connected nets of nodes that can be naturally adapted to unstructured meshes: the convolution in a GNN is performed by aggregating information from neighbouring nodes, thus overcoming the limitations imposed by the geometry in contrast with convolutional NN (CNN). Moreover, GNNs: i) show remarkable generalization capabilities as compared to standard network models \citep{sanchez2018graph}; ii) are differentiable; iii) provide the possibility of directly targeting the learning of the operator via discrete stencils \citep[see][]{shukla2022scalable}. Due to these features, this architecture has recently attracted attention in fluid mechanics. A review is available on the subject authored by \cite{lino2023current}, while examples are given by the works of \cite{toshev2023learning} or \cite{dupuy2023modeling}, where wall shear stress are modelled for LES simulations based on GNN. Here, we take inspiration from \cite{DSS}, where a GNN-based architecture incorporating permutation and translation invariance is combined with the statistical solver problem; \cite{DSS} proofed that the architecture -- referred to as deep statistical solver -- has some universal approximation properties, and it is capable of operator learning. 

\smallskip
In this contribution, GNN are combined with numerical simulations performed using finite elements method (FEM). The GNN-FEM interface allows the use of NN predictions in post-processing FEM analysis since it provides a two-way coupling between NN and FEM environments. However -- as already mentioned -- the volume of data can represent a bottleneck if the available amount is insufficient. Moreover, also quality of data impacts on the prediction properties in pure data-driven modelling \citep{bucci2021leveraging}, in particular when data at hand are not sufficient in representing correctly the distribution of the overall dataset. Inconsistent or unbalanced data distributions may lead the model to being trained on subsets of data that do not adequately represent the underneath physics of the problem. Thus, as second, fundamental ingredient we adopt active learning, aimed to increase the dissimilarity between data points and ensure that the model distills all discriminant features necessary to perform the required task effectively. At the best of our knowledge, this application is among the first in fluid mechanics where selection of data is performed by applying an active learning criterion. Readers are directed to \cite{ren2021survey} for a comprehensive overview of active learning methods in deep learning. We follow the work by \cite{input_similarity} where scalar products of gradients associated with the update of the model weights of the GNN are considered as similarity metric between samples; the chosen criterion allows to dynamically increase the training dataset by introducing data that promote diversity in the dataset. %The diversity criterion is not the only useful metric for conducting active learning; alternative approaches may incorporate prediction uncertainties.

\smallskip
The remainder of the paper is organised as follows. We state the fluid mechanics problem in Sec.~\ref{sec:fluido} and Sec.~\ref{sec:design_of_experiment}, where we discuss the set of equations used for the numerical simulations, the flow cases and the computational setup (Sec.~\ref{subsec:computational_domain}). The theoretical background is complemented by a discussion of the data-driven modelling based on GNN in Sec.~\ref{sec:methodology}, and how this architecture is coupled with numerical simulations resolved on unstructured meshes. The second part of the paper focuses on the results (Sec.~\ref{sec:results}): GNN models are trained on datasets composed by the mean flows (input) and the Reynolds stress (target); the baseline is provided by the closure problem in cylinder flows at Reynolds numbers $50 \leq Re \leq 150$ (Sec.~\ref{subsec:base_case}). Generalization properties are assessed using $4$ different benchmarks defined in Sec.~\ref{subsec:test_cases}, not included in the training sets. This baseline is thus compared with models trained with different strategies of data selection. First, we consider cases where data augmentation is performed using simulations of flows past bluff bodies of random geometry (Sec.~\ref{subsec:random_shape}); second, active learning is introduced in Sec.~\ref{subsubsec:active_learning}. The paper finalises with conclusions in Sec.~\ref{sec:conclusions}.
\section{Governing equations}
\label{sec:fluido} 
In this work, we will focus on incompressible two-dimensional (2D) fluid flows developing past bluff bodies in unsteady flow regimes. We consider the Navier--Stokes (NS) equations for incompressible flows. By denoting the spatial Cartesian coordinates as $\mathbf{x}=(x,y)$, the dynamics of the velocity field ${\vdir}(\mathbf{x},t)$ and pressure field $p(\mathbf{x}, t)$ are governed by 
\begin{subequations}
    \begin{align}
        \dfrac{\partial \vdir}{\partial t} +\left(\vdir\cdot\nabla\right) \vdir &= -\nabla\pdir +\invRe\nabla^2\vdir
        \\
        \nabla\cdot\vdir &= 0.
    \end{align}\label{eq:unsteadyNS}
\end{subequations}
Eq.s~\ref{eq:unsteadyNS} are made dimensionless using the characteristic length scale $D$ (\textit{e.g.} the cylinder diameter), the velocity of the incoming uniform flow $U_{\infty}$ as the reference velocity and $\rho U_{\infty}^2$ as the reference pressure. Based on these quantities, the Reynolds number is defined as $Re = \nicefrac{U_{\infty} D}{\nu}$, where $\nu$ is the kinematic viscosity.

\smallskip
As baseline equations for the data assimilation process, we will consider the Reynolds-averaged Navier--Stokes (RANS) equations, the time averaged formulation of the NS Equations \cite{foures}. Notice that, although typically used for turbulent flows,  the RANS approach is well suited also for unsteady flow, for instance for the determination of unsteady periodic flow dynamics through weakly nonlinear analyses \cite{SIPP_LEBEDEV_2007,manticlugo}. Thus, we introduce the Reynolds decomposition
\begin{equation}
\vdir(\mathbf{x},t) = \mean\vdir(\mathbf{x}) + \fluct\vdir(\mathbf{x},t),\label{eq:reydeco} 
\end{equation}
where the vector field $\vdir=(u,v)^T$ is rewritten as sum of the time-averaged velocity vector field $\mean{\vdir}=(\mean{u},\mean{v})^T$ and the fluctuation velocity vector field $\fluct\vdir=({\fluct u},{\fluct v})^T$. Formally, any unsteady flow can be described through this decomposition. Plugging the Reynolds decomposition in Eq.~\ref{eq:reydeco} into the Navier--Stokes equations (Eq.~\ref{eq:unsteadyNS}) and time averaging, we get the system of equations
\begin{subequations}
    \begin{align}
        \mean\vdir\cdot\nabla\mean\vdir + \nabla\mean\pdir - \invRe\nabla^2\mean\vdir &= \ff \\
        \nabla\cdot\mean\vdir &= 0,
    \end{align}\label{Eq:RANS}
\end{subequations}
where $\mean{p}$ is the average pressure field. In this framework, the forcing term $\ff$, acting as a closure term, is identified as the Reynolds stress tensor. In principle, it can be directly computed when data are available as
\begin{equation}
\ff= -\nabla \cdot (\overline{\fluct\vdir \fluct\vdir}).\label{eq:forcing}
\end{equation}
In practice, computing the forcing term in this way requires the solution of DNS or time-resolved experimental measurements, as the fluctuations do not directly depend on the meanflow. This issue is commonly referred to as the closure problem and, for turbulent flows, a number of approximated models based on the Boussinesq hypothesis such as $k-\epsilon$ or $k-\omega$ \citep{wilcox1998turbulence} or more complex models such as explicit algebraic Reynolds stress model \citep{wallin2000explicit} or differential Reynolds stress models \citep{cecora2015differential} have been introduced. Here, we will rather focus on a data-assimilation approach in the footprints of \cite{foures} and \cite{patel2024turbulence}, where assumptions such as the Boussinesq one are not used. In this context, we aim at approximating the closure term as a forcing term of an optimization process without relying on an adjoint-based loop. A model based on Graph Neural Networks (GNN) is trained as a surrogate model where the output is the forcing term (Eq.~\ref{eq:forcing}) and the input is given by the meanflow $\mean\vdir$. 
\section{Design of experiment and numerical setup} \label{sec:design_of_experiment} 
The main goal of the present work is to develop a GNN-based surrogate model emulating the closure term $\ff$ given as the target output and the meanflow as input. Thus, in order to train the GNN, a fluid dynamics dataset is built. In this section, we consider bluff bodies geometries at different Reynolds numbers, such that a comprehensive dataset is obtained to handle a larger variety of flow scenarios. In the following, the reference case -- \textit{i.e.} the cylinder flow -- is introduced in Sec.~\ref{subsec:clyflow}, while the flows past random geometries are discussed in Sec.~\ref{subsec:random_shape_generation}. We briefly report on the numerical simulations performed for this scope in Sec.~\ref{subsec:computational_domain} and more deeply in App.~\ref{appA:numerical_sim_details}.

\subsection{Cylinder flow}
\label{subsec:clyflow} 
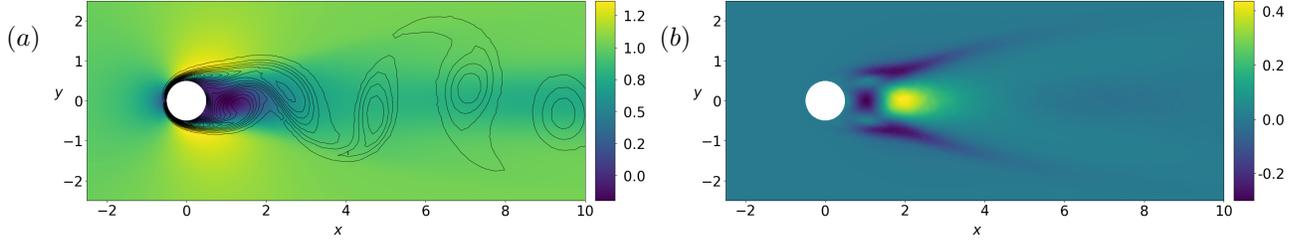
\begin{figure}[t]
\includegraphics[width=1\textwidth]{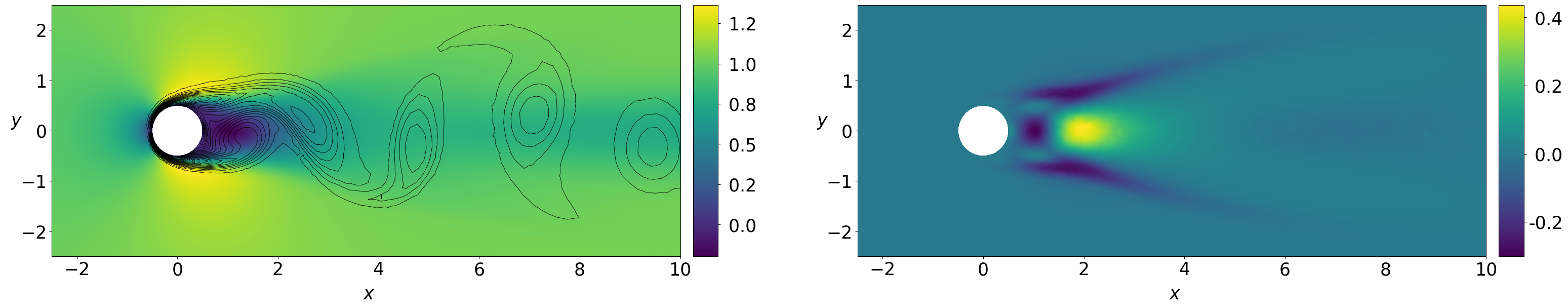}
\caption{$(a)$ Stream-wise component of the meanflow $\mean \vdir$ and vorticity isolines $\omega = \nabla \times \vdir$ for the flow past a cylinder at $Re = 150$. $(b)$ For the same case, the stream-wise component of the closure term $\mathbf{f}$ is shown. In both cases, only a portion of the domain is shown.}\label{fig:mean_vort}
\begin{picture}(0,0)
\put(-250,85){$(a)$}
\put( -03,85){$(b)$} 
\end{picture}
\end{figure}
The unsteady wake developing past a cylinder is a well documented benchmark in fluid dynamics literature, and it is often found as reference case in works of data-assimilation for the assessment of novel techniques; it exhibits steady behaviour until a critical Reynolds number of $Re_{c} \cong 46.7$, when a supercritical Hopf bifurcation occurs \citep{Provansal_Mathis_Boyer_1987, giannetti2007structural}.
At higher Reynolds numbers, the baseflow becomes an unstable solution and the unsteady flow develops into a limit cycle known as von Karman street (Fig.~\ref{fig:mean_vort}$a$). This behaviour can be observed until $Re = 150$, for two-dimensional (2D) cases. Beyond this point, alongside periodic vortex formation, irregular velocity fluctuations begin to emerge \citep{roshko}; note that transition to turbulence occurs when three dimensional cases are considered, which is beyond the scope of the present work. Instead, this work aims at a quantitative analysis of GNN model emulators trained in low-data limits leveraging active learning processes; as such, we will benchmark the proposed algorithm on simpler 2D scenarios exhibiting the limit cycle behaviour, in the range $50 \le Re \le 150$; we perform numerical simulations at different $Re$ numbers in the mentioned range, with $\Delta Re = 10$, in order to collect the necessary data to train the GNN model and compose the dataset. The numerical simulations are resolved in time, while the meanflow $\mean \vdir$ and the forcing term $\ff$ (Eq.~\ref{eq:forcing}) are computed by averaging on-the-fly until convergence. As an example of this data, Fig.~\ref{fig:mean_vort}$(a)$ illustrates the stream-wise component of the meanflow $\mean \vdir$ alongside the vorticity isolines $\omega = \nabla \times \vdir$, while Fig.~\ref{fig:mean_vort}$(b)$ shows the stream-wise component of the closure term $\ff$ for $Re=150$.

\subsection{Flow around random shapes}
\label{subsec:random_shape_generation} 
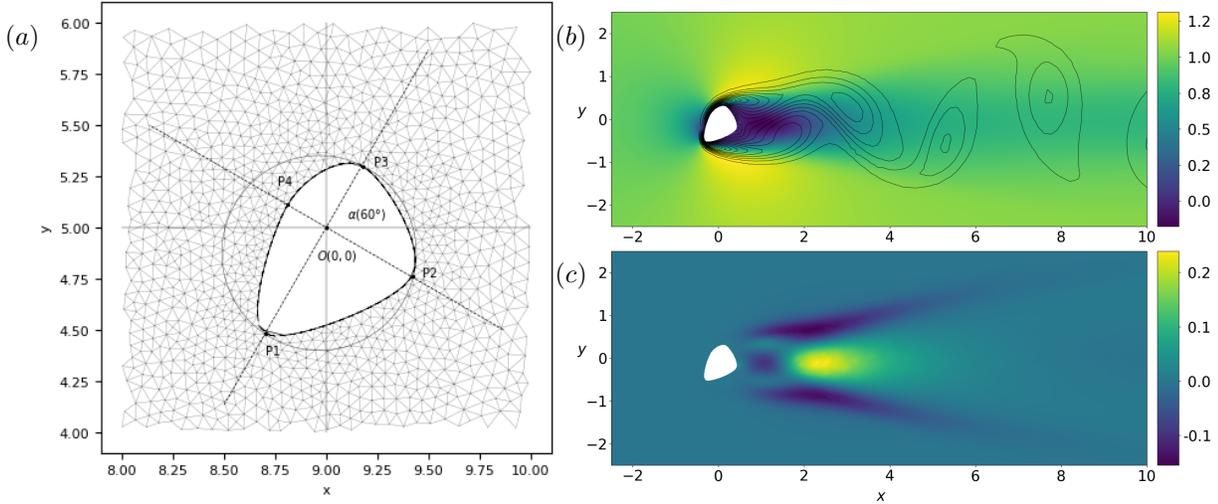
\begin{figure}[t]
\centering
\includegraphics[width=0.43\textwidth]{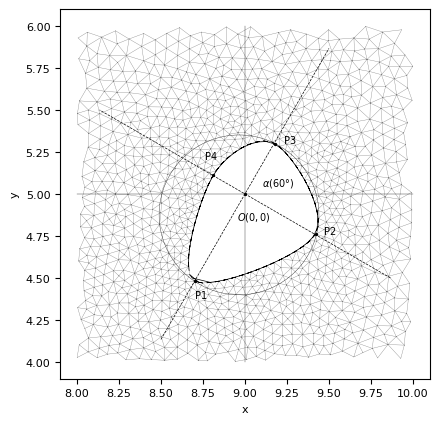}
\includegraphics[width=0.52\textwidth]{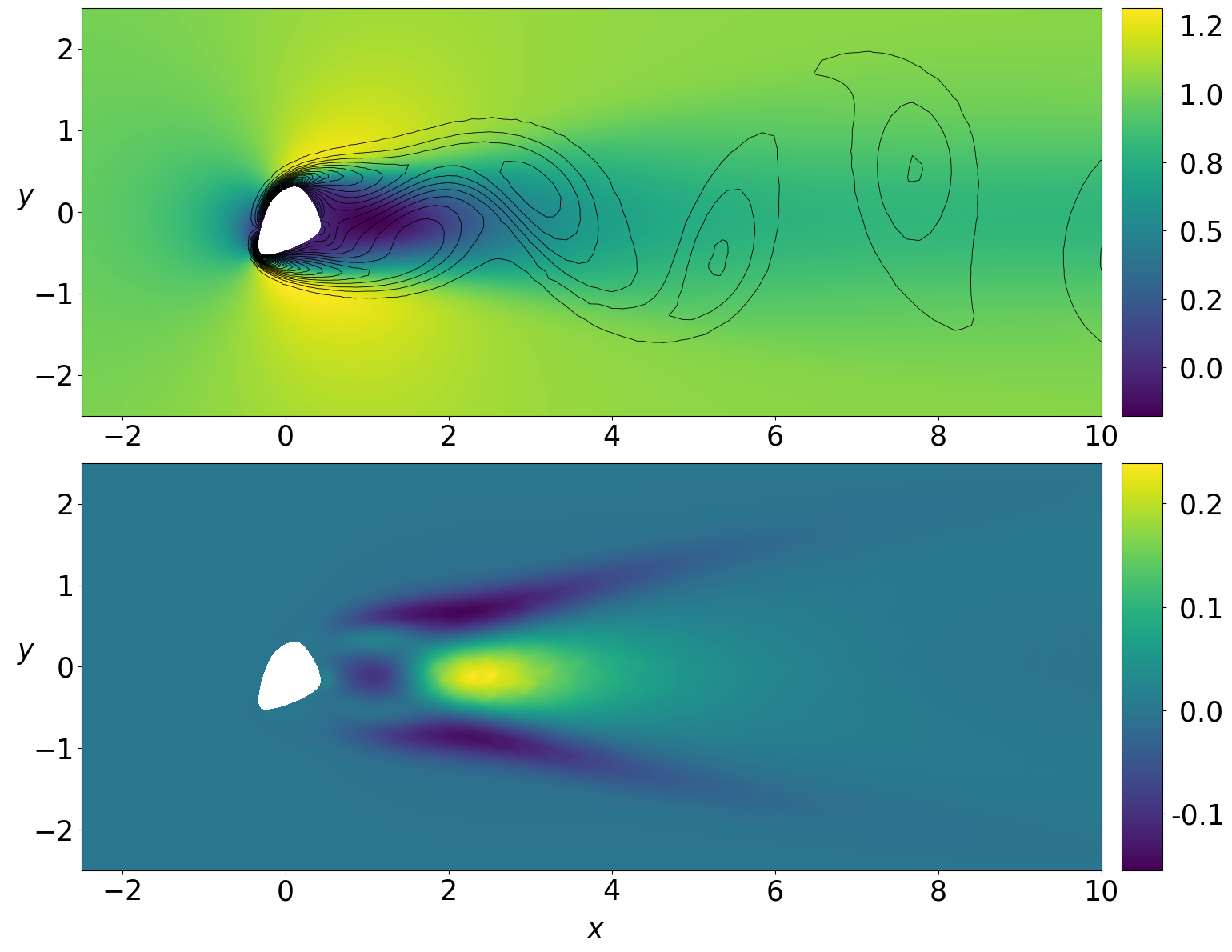}
\begin{picture}(0,0)
\put( -460,175){$(a)$} 
\put( -252,175){$(b)$} 
\put( -252,85){$(c)$} 
\end{picture}
\caption{Flow past a bluff body with a randomly generated shape at $Re=90$. $(a)$ Geometry of the bluff body with $\alpha = 60\degree$ and unitary diameter of the circumscribed circle around the body. The resulting meanflow and the vorticity isolines $\omega = \nabla \times \vdir$ are shown in $(b)$, while $(c)$ shows the forcing stress term in a portion of the domain.}
\label{fig:random_shape}
\end{figure}
As mentioned previously, the dataset can be enriched by including different geometries; thus, alongside with the flow past a cylinder, we include in the dataset flow fields around obstacles of random shapes. In principle, we should determine the critical Reynolds number ($Re_c$) at which the flow around each of the random geometries under analysis develops into unstable baseflows. This would require an extensive campaign of simulations and a comprehensive stability analysis, along the footprints of the recent works by \cite{chiarini2021rectangulat, chiarini2022symmetric}. Therefore, we adopt a more pragmatic approach by retaining the geometries that in the range  $50 \le Re \le 150$ develop unsteady flows.

\smallskip
Regarding the definition of the objects geometries, each shape is defined by a set of splines connected at the location of $4$ control points, assuring a $C1$ continuity between them. The control points (P1 to P4 in Fig.~\ref{fig:random_shape}$a$) are located along two orthogonal axes rotated of an angle $\alpha$ with respect to the Cartesian coordinates system of reference. Their distance from the origin $O(0,0)$ is defined in the range $|s| \in [0.1, 0.6]$, with a discrete step size of $\Delta s = 0.1$. An additional degree of freedom is added for controlling the angle $\alpha \in [0^{\circ}, 90^{\circ}]$, with a discrete step size of $\Delta \alpha = 30 ^{\circ}$.

\smallskip
Finally, the random shapes generation script prevents the repetition of the same set of degrees of freedom defining the shape such that each geometry appears only once in the final dataset. Note that the centre of the circumscribed circle of each random shape does not align necessarily with the centre of the coordinate system. This is not an accidental feature: in the context of NN training, this strategy corresponds to introducing some positional noise in the training dataset, thus promoting a more robust and prone-to-generalize learning of the underlying data. Such variability in the training data further contributes to reduce the risk of overfitting and enhances the NN ability to generalize its predictions to new, unseen geometries.

\subsection{Numerical setup}
\label{subsec:computational_domain}
\begin{figure}[t]
\centering
\includegraphics[width=.8\textwidth]{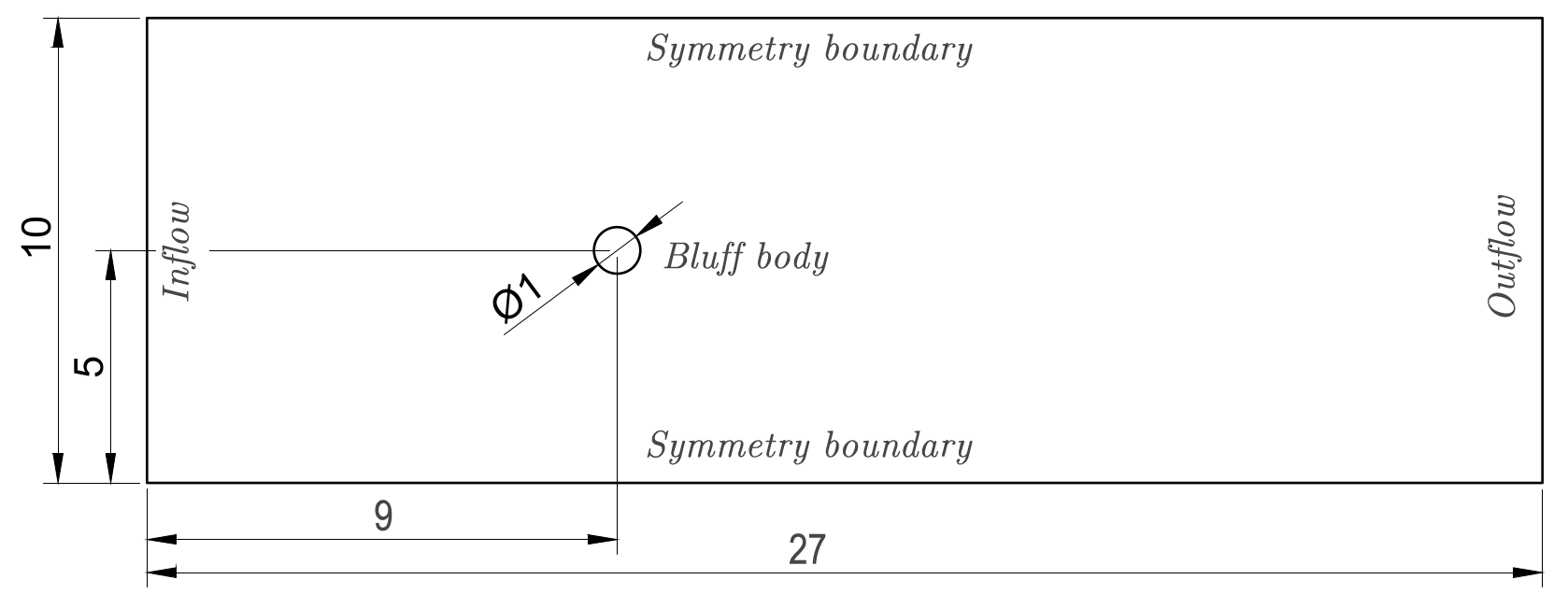}
\caption{Sketch of the computational domain geometry. The diameter of the circumscribed circle of the bluff body, the height and length of the domain are given in non-dimensional units.}
\label{fig:case_study}
\end{figure}
In our numerical setup, we use as a characteristic dimension the diameter $D$ of the circumscribed circle of each of the generated bluff bodies. Based on this dimension, the domain extension is $L_x=27$ in the stream-wise direction and $L_y=10$ in the transverse direction; the centre of the system coordinates is placed at $O(0, 0)$, at a distance $\Delta x=9$ from the inlet in the stream-wise direction and $\Delta y=5$ from the symmetry boundaries.

\smallskip
The flow evolves from the inlet, where a dimensionless uniform velocity $\vdir=(1,0)^T$ is imposed, normalized with respect to the reference velocity $U_{\infty}$. Following the study of \cite{foures}, we introduce the following set of boundary conditions
\begin{eqnarray}
\begin{cases}
\begin{array}{r l}
u = 1, \hspace{5pt} v = 0 &\text{at the inlet,} \\
u = 0, \hspace{5pt} v = 0 &\text{on the cylinder surface,}\\
\partial_y u = 0, \hspace{5pt} v = 0 &\text{on symmetry boundaries,}\\
\invRe\partial_x u - p = 0, \hspace{5pt} \partial_x v = 0 &\text{at the outlet}.\\
\end{array}
\end{cases}
\end{eqnarray}
All the numerical simulations are initiated with null flow fields at $t=0$. The required statistics, \textit{i.e.} meanflow and second order moments, are computed on-the-fly during the simulations. The final time $T$ of the simulation is determined by a convergence criterion based on the L2-norm of the difference between two subsequent meanflows below a threshold of $10^{-8}$. 

\smallskip
Finite element method (FEM) is used for the spatial discretization, based on the \verb}FEniCS python} library \citep{alnaes2015fenics}; time marching is performed by second order backward differentiation formula (BFD). The meshes used for the numerical simulations are refined near the obstacle and in the wake region to best capture the flow dynamics. Depending on the shape of the obstacle, our meshes count approximately $13500$ nodes for the whole domain; more details on the adopted numerical schemes and convergence of simulations are provided in \ref{appA:numerical_sim_details}.
\section{Methodology}\label{sec:methodology}
In this section, we aim at providing a concise overview of some of the fundamental features that characterize a GNN. An exhaustive and general description of the main features of this architecture can be found in \cite{hamilton2020graph}. Here, we focus on the main process on which a GNN is based: the diffusion of the information extracted from nodes and connections across the whole network using the {message passing} ({MP}) algorithm; Sec.~\ref{subsec:intro_method} is mainly dedicated to {MP}, while Sec.~\ref{subsec:data_struct} reports on the data structure. The training process is described in Sec.~\ref{subsec:training_process}.

\subsection{Neural Network architecture}\label{subsec:intro_method}
\begin{figure}[t]
\center
\includegraphics[width=0.7\textwidth]{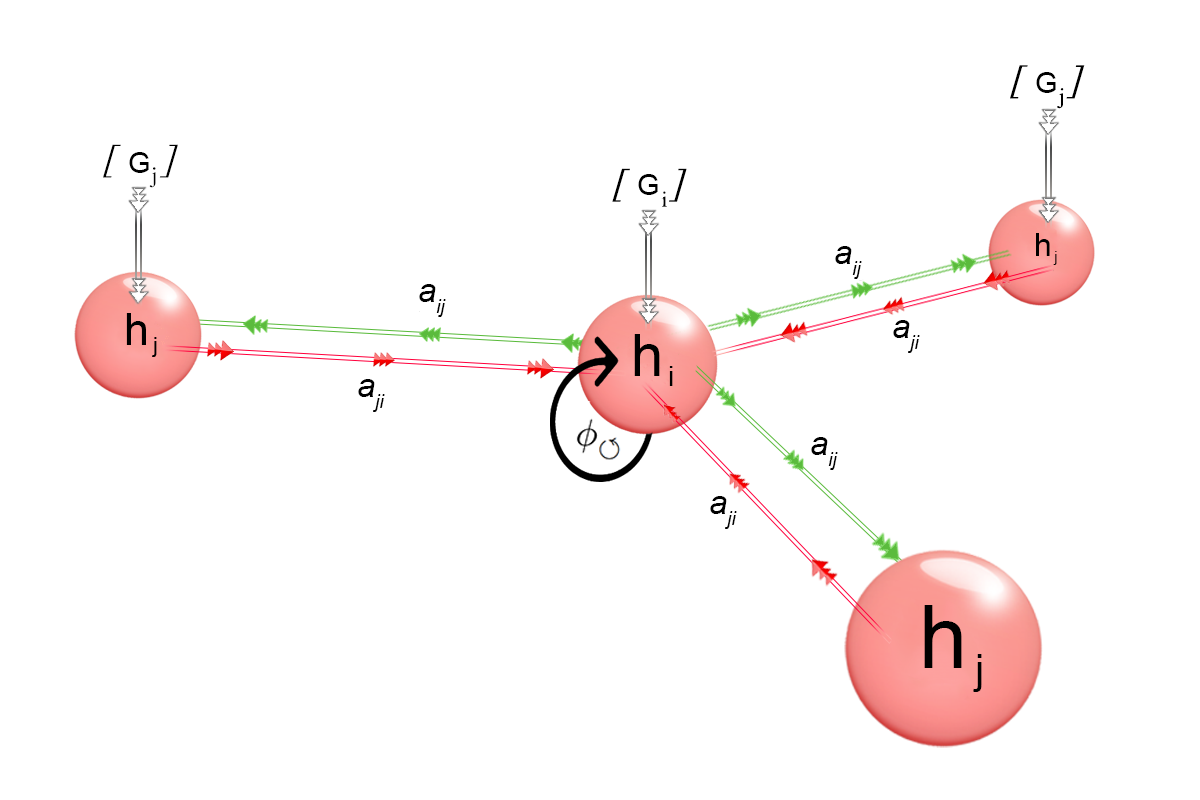}
\caption{Pictorial sketch of the Graph Neural Network message-passing process: given the $i$-th node, $\mathbf{h}_i$ is the embedded state defined on $i$; ${a}_{ij}$ are the directed connections between the nodes; $\textbf{G}_i$ represents the external inputs injected in each node (see Sec.~\ref{subsec:intro_method} for details).}\label{fig:GNN_structure}
\end{figure}

Message passing can be divided into three fundamental steps, which we describe in the following as applied to our specific case.  A pictorial sketch summarising this process is shown in Fig.~\ref{fig:GNN_structure}.
\begin{enumerate}
\item \textbf{Message creation} -- on each node $i$, an embedded state associated with the vector $\textbf{h}_i$ is created. It is initialized with a zero state and it will embed information as long as the message passing proceeds. The dimension $d_h$ of the vector $\textbf{h}_i$ is constant across all nodes and it is a fundamental model hyperparameter that defines, along with the number of update layers $k$, the {expressivity} \citep{IngoRaslan2007expressivity} of the GNN, \textit{i.e.} the ability of the model to represent functions of a certain complexity. The embedded state does not have a direct physical meaning.
\item \textbf{Message propagation} -- in this step, the information is propagated. To capture the convective/diffusive dynamics of the underneath system, \textit{i.e.} reproducing a model for the RANS equations, we transmit the message in both directions: from each node $i$ to its neighbour nodes $j$ and vice versa, from the nodes $j$ to the starting node $i$ (see Fig.~\ref{fig:GNN_structure} for a sketch). Mathematically, for each couple of connected nodes $i$ and $j$, we can write the following relation
\begin{equation}
\boldsymbol{\phi}_{i, j}^{(k)} = \zeta^{(k)} \left(\textbf{h}_{i}^{(k-1)}, \textbf{a}_{ij}, \textbf{h}_{j}^{(k-1)}\right).
\label{eq:phi}
\end{equation}
In what follows, the subscript will indicate the node index, while the superscript the update index. The vector $\textbf{h}^{(k-1)}_{i}$ is the embedded state from the previous layer $k-1$, $\textbf{a}_{ij}$ denotes the directed connections between the node $i$ and the nodes $j$, while $\zeta^{(k)}$ is a generic differentiable operator such as Multi-Layer Perceptron (MLP), see for instance \cite{goodfellow2016deep}.
\item \textbf{Message aggregation} -- finally, on each node $i$, these collected information are aggregated, providing an updated embedded state $\textbf{h}^{(k)}_{i}$ for each of them
\begin{equation}
\textbf{h}^{(k)}_i = \textbf{h}^{(k-1)}_i + \alpha \Psi^{(k)} \left( \textbf{h}^{(k-1)}_i, \textbf{G}_i,\boldsymbol{\phi}^{(k)}_{i,\rightarrow}, \boldsymbol{\phi}^{(k)}_{i,\leftarrow}, \boldsymbol{\phi}^{(k)}_{i,\circlearrowright}\right),
\end{equation}
where the inputs are the meanflow and the Reynolds numbers, contained in the vector $\textbf{G}_i = \{\mean\vdir, Re\}$ and are fed at each update; ${\boldsymbol{\phi}^{(k)}_{i,\rightarrow} = \frac{1}{N_i} \sum_{j\in N_i} \boldsymbol{\phi}^{(k)}_{i, j}}$ is the message that the node $i$ sends to its {$N_i$} neighbours, averaged over the number of neighbours; ${\boldsymbol{\phi}^{(k)}_{i,\leftarrow} = \frac{1}{N_i} \sum_{j\in N_i} \boldsymbol{\phi}^{(k)}_{i,j}}$ is the message that the node $i$ receives from its {$N_i$} neighbours, averaged over the number of neighbours; ${\boldsymbol{\phi}^{(k)}_{i,\circlearrowright} = \boldsymbol{\phi}^{(k)}_{i, i}}$ the message that the node $i$ sends to itself to avoid loss of information as the process advances. $\Psi^{(k)}$ is a generic differentiable operator that can be approximated by an MLP also in this case; $\alpha$ is a relaxation coefficient, included in the model hyperparameters, that allows to scale each update of the embedded states with the previous one during the message passing loop. Note that the use of the mean operator in ${\boldsymbol{\phi}^{(k)}_{i,\rightarrow}}$ and ${\boldsymbol{\phi}^{(k)}_{i,\leftarrow}}$ allows us to collect, for each of the $i$-th embedded states, the information propagating from a number of neighbours that can differ for each of the $i$-th nodes. Nonetheless, different permutation-invariant functions can be introduced in this step (maximum, sum, concatenation...).
\end{enumerate}
At the end of the message passing process, the information on each node has been handled as described and thus carries with it information from any other point of the graph including their relative distances. In order to allow the message on each node to reach any other node of the mesh, our intuition is that -- in principle -- the number of updates $k$ should cover the longest geodesic path that can be defined on the mesh \citep{DSS}; in practice, the number of updates in the present work is fixed and optimized using genetic algorithms (see \cite{optuna_2019} and {\ref{appB:GNN_optimization}} for details). The latest embedded state is projected back to a physical state using a decoder, once again approximated by a differentiable function that in our case is an MLP.

\subsection{Data structuring}\label{subsec:data_struct}
The application of GNN combined with unstructured meshes revolves around the idea of representing mesh data as a graph. In this representation, each node of the mesh is coupled one-to-one to a node in the GNN. In order to achieve this overlap, we structure the data in a tensor form that can be handled as a graph while preserving the adjacency property from the mesh. We build, for each case:
\begin{itemize}
    \item \textbf{Node Feature Matrix} $\mathbf{A}$: This matrix $\mathbf{A} \in \mathbb{R}^{n_i \times d_h}$ represents the features of each node in the mesh. Here, $n_i$ is the number of nodes of the mesh, and $d_h$ is the dimension of the embedded state in which we project the FEM quantities; each node is assigned with a unique index.
    \item \textbf{Edge Index Matrix} $\mathbf{C}$: The matrix $\mathbf{C} \in \mathbb{R}^{c \times 2}$ defines the edges in the mesh. Here $c$ represents the number of edges of the mesh, and each row contains a pair of indices denoting connected nodes.
    \item \textbf{Edge Attribute Matrix} $\mathbf{D}$: The matrix $\mathbf{D} \in \mathbb{R}^{c \times 2}$, contains the computed distances in $x$ and $y$ directions between the connected nodes. This distances include their signs, and $\mathbf{D}$ shares the same row indices $c$ as $\mathbf{C}$.
\end{itemize}
The matrices are defined as follows:
\[
\textbf{A}_{n_i,d_h} =
\begin{bmatrix}
a_{1,1} & a_{1,2} & \cdots & a_{1,d_h} \\
a_{2,1} & a_{2,2} & \cdots & a_{2,d_h} \\
\vdots  & \vdots  & \ddots & \vdots  \\
a_{n_i,1} & a_{n_i,2} & \cdots & a_{n_i,d_h}
\end{bmatrix},
\]
\\
\[
\textbf{C}_{c,2} =
\begin{bmatrix}
i_1 & j_1\\
i_2 & j_2\\
\vdots  & \vdots \\
i_c & j_c
\end{bmatrix},
\qquad
\textbf{D}_{c,2} =
\begin{bmatrix}
x_{i_1} - x_{j_1} & y_{i_1} - y_{j_1}\\
x_{i_2} - x_{j_2} & y_{i_2} - y_{j_2}\\
\vdots  & \vdots \\
x_{i_c} - x_{j_c} & y_{i_c} - y_{j_c}\\
\end{bmatrix}.
\]
Each column of the matrix $\mathbf{A} \in \mathbb{R}^{n_i \times d_h}$ is assigned as a feature vector to a single neuron on the MLPs structure employed in the MP step of the GNN, namely $\zeta$ (see Sec.~\ref{subsec:intro_method}). These feature vectors are set to zero at the beginning of the training process (see Sec.~\ref{subsec:training_process}). The MLP structure is defined by the dimension $d_h$ of the embedded state, rather than by the number of features per node. This ensures that the MLP architecture remains unaltered regardless of the number of nodes in the mesh. Indeed, changing the number of nodes in the mesh only affects the row index $n_i$ in matrix $\mathbf{A} \in \mathbb{R}^{n_i \times d_h}$, the number of connections $c$ in matrix $\mathbf{C} \in \mathbb{R}^{c \times 2}$, and the distances in $\mathbf{D} \in \mathbb{R}^{c \times 2}$ but keeps the MLP structure intact, as it is defined on the dimension $d_h$ of the embedded state. This ensures that the same MLP architecture can be used for different FEM simulations, regardless of varying geometries or node numbers.
Moreover, the GNN is not informed with absolute spatial coordinates as inputs, meaning it does not learn any specific mapping between spatial positions and the target output. This ensures that the GNN maintains invariance with respect to the position of the bluff body and mesh refinement, enhancing its generalization capability across different simulation setups.
The ability of interacting with unstructured meshes, learning from various geometries, and reconstructing forcing fields on different geometries is rather straightforward when considering GNN architectures as compared with standard techniques such as CNN, thanks to their versatility. Finally, the model is equipped with a FEM-GNN interface that enables the mapping of numerical results from the GNN to a vector field computed or stored in the FEM environment.

\subsection{Training algorithm} \label{subsec:training_process}
\begin{figure}[t]
    \centering
    \includegraphics[width=.825\textwidth]{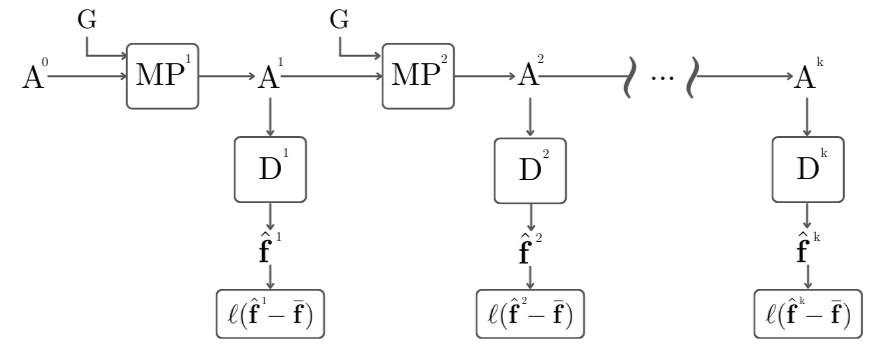}
    \caption{Figure adapted from \cite{DSS} showing the overall framework of the GNN training procedure. ${MP}^k$ are the message passing algorithms; $\textbf{D}^k$ are the $k$ decoders trainable MLPs; $\textbf{A}^k$ are the $k$ matrices containing the embedded states from each node; $\textbf{G}$ is the vector containing the input injected in the GNN. Refer to Sec.~\ref{subsec:training_process} for details.}
    \label{fig:dss_overall}
\end{figure}

The training framework adopted for the presented GNN is described below, with reference to Fig.~\ref{fig:dss_overall}. The starting point is $\textbf{A}^{0}$ which represents the matrix composed by all the embedded states defined on the nodes, initialized at zero state. Along with the externally injected quantities ($\textbf{G}$, the meanflow and the Reynolds number $Re$ in our case), it is provided to the message passing process ${MP}^{1}$ (see Sec.~\ref{subsec:intro_method}). This latter will output an updated version of the embedded state tensor, $\textbf{A}^{1}$ which will pass through a decoder, $\textbf{D}^{1}$, a multi-layer perceptron (MLP) trainable function in charge of reconstructing a meaningful physical state ($\hat{\ff}^{1}$) from the embedded state which, in our case, is the closure term of the RANS equations. Finally, this physical state is compared with the ground truth that comes from the DNS ($\bar \ff$) via a loss function $\ell =(\hat{\ff} - \bar \ff)^2$. The entire process is repeated in each layer until the final layer $k$ is reached. The last prediction ($\hat{\ff}^k$) represents the output of the entire algorithm. Following the procedure framework in \cite{DSS}, all the intermediate loss values are considered, in order to robustify the learning process, through a global loss function:
\begin{equation}
    L=\sum_{k=1}^{\bar{k}} \gamma^{\bar{k}-k} \left( \frac{1}{n_i} \sum_{i=1}^{n_i} \ell_i \right), \label{eq:cost}
\end{equation}
in which $n_i$ is the number of nodes, $\bar{k}$ is the number of update layers and $\bf{\gamma}$ is an hyperparameter used to control the importance of the contributions associated with each update of the embedded states. The relative importance of the message is weighted by the exponential term $\gamma^{\bar{k}-k}$, with $\gamma<1$; thus, the latest updates $k$ will be the richest in information and have the highest importance in the process. In {\ref{appB:GNN_optimization}}, the list of hyperparameters and their values after optimization are reported.
\section{Results}
\label{sec:results} 
The present section is dedicated to the presentation of the results. We introduce a proof-of-concept of the approach in Sec.~\ref{subsec:base_case}; in this first analysis, we train the GNN with a dataset composed of data obtained from simulations of the flow past a cylinder at different Reynolds numbers. The aim is to verify the robustness of the approach with respect to unseen conditions, using the benchmark cases in Sec.~\ref{subsec:test_cases}. In Sec. \ref{subsec:base_case}, we focus on the appropriate choice of the dataset and how the selection of training data can impact the generalization capabilities of the GNN. In Sec.~\ref{subsec:random_shape}, we include in the dataset flow fields from simulations of wakes past randomly generated bluff bodies (details of shape generation in Sec.~\ref{subsec:random_shape_generation}); the effects of dataset augmentation and dataset expansion are discussed in Sec.~\ref{subsubsec:data_augmentation} and Sec.~\ref{subsubsec:dataset_expansion}, respectively, while data selection by active learning criteria is discussed in Sec.~\ref{subsubsec:active_learning}. A quantitative comparative analysis of these training approaches is summarized in Sec.~\ref{sec:comparison}; in the following, we introduce the $4$ reference cases used for all the comparisons discussed in this section.

\subsection{Test Cases}
\label{subsec:test_cases}
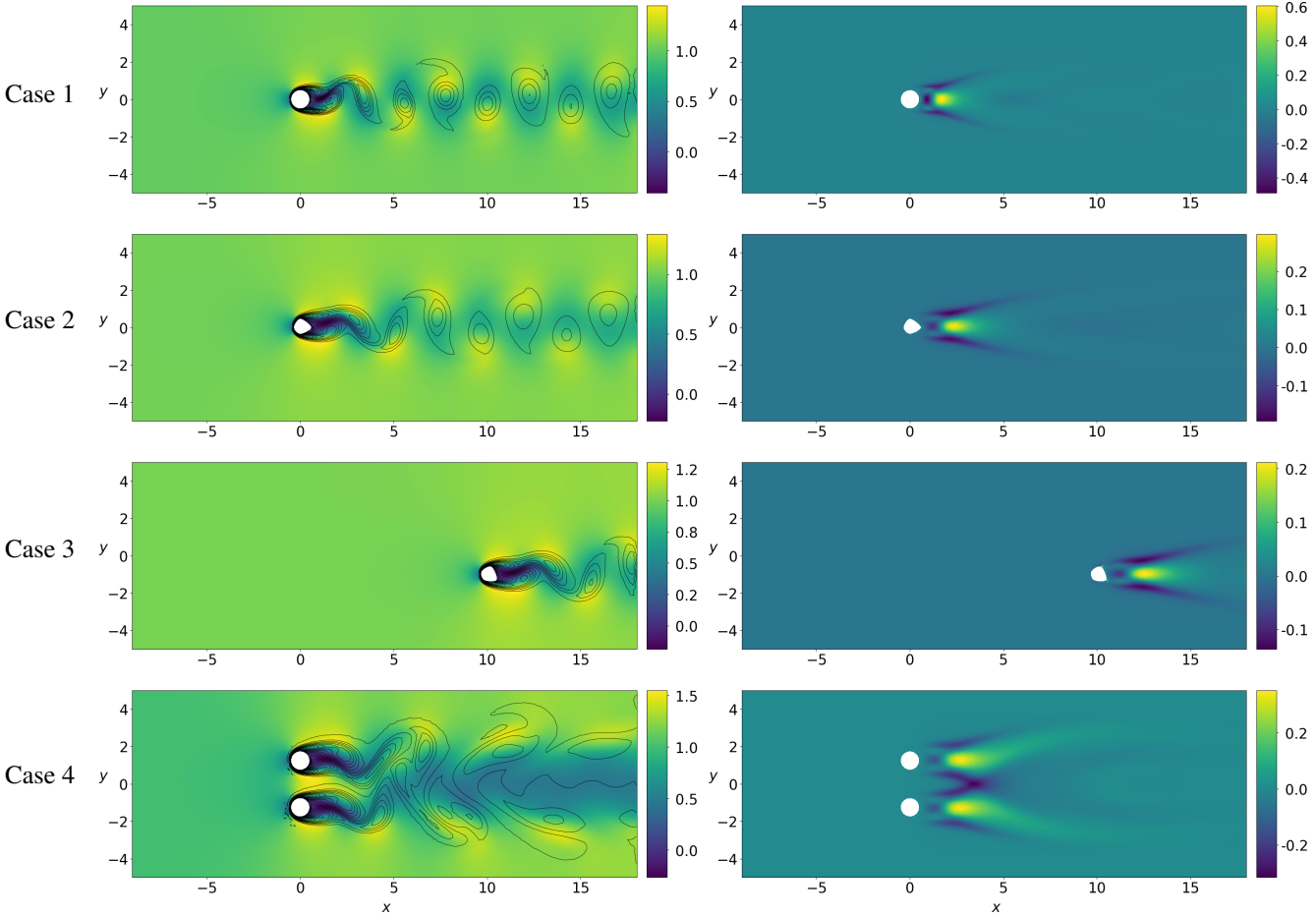
\begin{figure}[t!]
    \centering
    \includegraphics[width=1\textwidth]{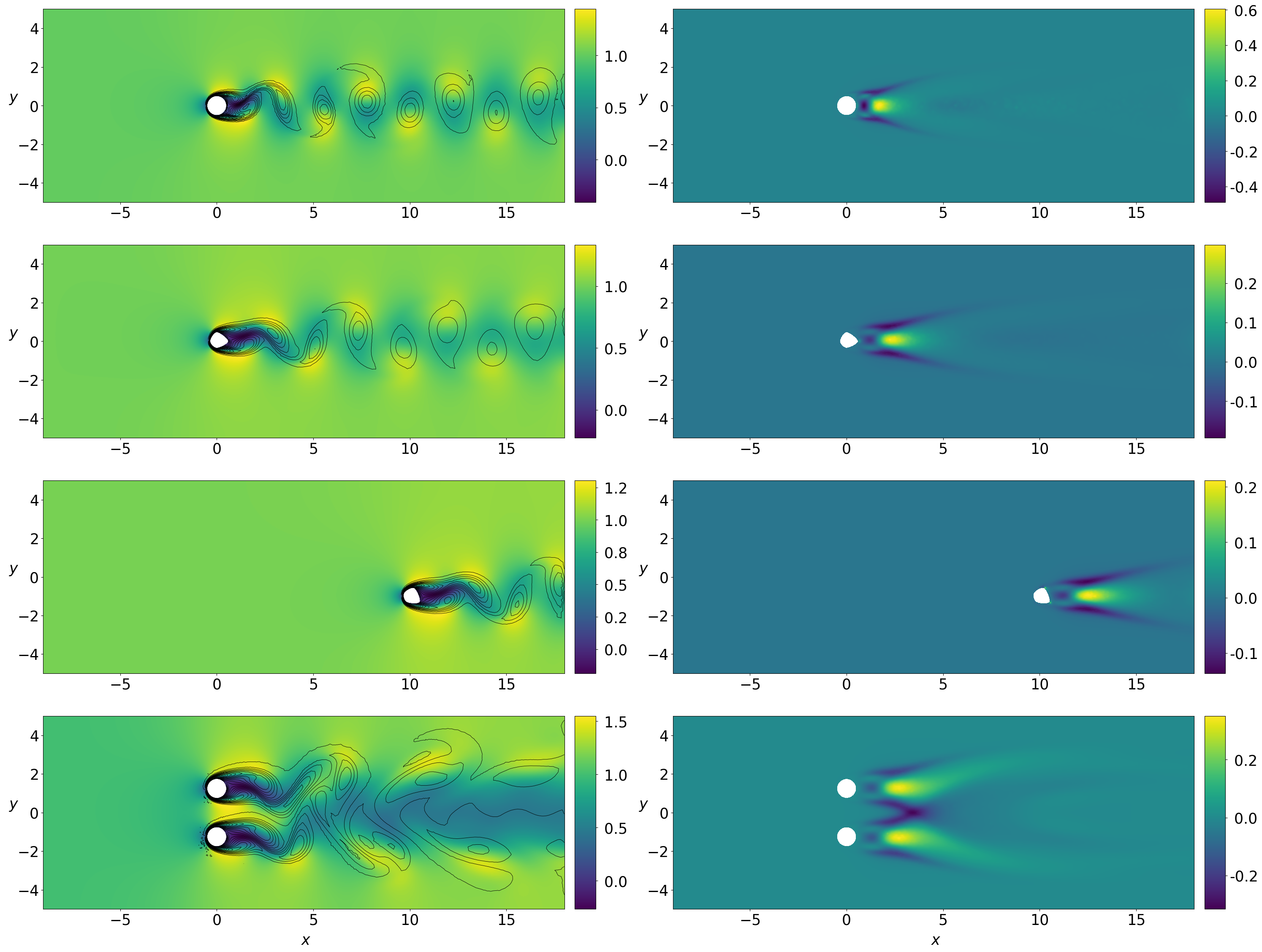}
    \begin{picture}(0,0)
        \put( -268, 325){Case~1}
        \put( -268, 238){Case~2} 
        \put( -268, 150){Case~3} 
        \put( -268, 63){Case~4} 
    \end{picture}
    \caption{Test cases used as a benchmark of the learning strategies discussed in Sec.~\ref{sec:results}. From the top raw to the bottom one: {Case~1}, cylinder flow at $Re = 200$; {Case~2}, flow past a random-shaped bluff body at $Re = 120$; {Case~3}, flow past a random-shaped bluff body shifted in the computational domain at $Re = 100$; {Case~4}, flow past a two side-by-side cylinders configuration at $Re = 90$. For each of the cases, the left column shows the stream-wise component of the meanflow $\mean \vdir$, along with the vorticity isolines $\omega = \nabla \times \vdir$, while the right column show the stream-wise component of the forcing stress $\ff$.}\label{fig:dns_test_cases}
\end{figure}

We introduce $4$ benchmarks to assess the performance of the GNN model. The test cases, labelled from Case $1$ to Case $4$, are designed to evaluate specific aspects of the GNN's prediction. We cover a broad spectrum of scenarios, including variability in Reynolds number and geometry, changes in the position with respect of the inlet and number of the obstacles.
\begin{enumerate}
\item \textbf{Case~1}: in this scenario, we consider the flow past a cylinder, where the Reynolds number is increased to $Re = 200$. This exceeds the training dataset interval, ranging in $50 \le Re \le 150$, and it is used to test the model's capability to extrapolate beyond the training data.
\item \textbf{Case~2}: here, we introduce as a test case the flow past a bluff body of random shape not included in the training dataset at $Re = 120$, in order to assess the model generalization capability to unseen shapes.
\item \textbf{Case~3}: this case involves data from simulation of a flow past a bluff body of random shape, not present in the training dataset, at $Re = 100$. The obstacle is positioned downstream of the reference position used in the dataset.
\item \textbf{Case~4}: this test considers a two side-by-side cylinders configuration; two bluff body obstacles are present in the flow, in contrast to the single obstacle used for the training. The Reynolds number is $Re=90$ and the aim is to test geometries inducing different dynamics, here the one stemming from multiple interacting bluff bodies.
\end{enumerate}
Fig.~\ref{fig:dns_test_cases} shows the streamwise component of the meanflow velocity $\mean \vdir$, the isolines of the vorticity $\omega = \nabla \times \vdir$ (left column) and streamwise component of the forcing stress $\ff$ (right column) computed using numerical simulations for the $4$ benchmarks Case~$1$-$4$.

\smallskip
Two metrics will be used to evaluate the performance of the GNN. The first one will be a comparison between the ground truth $\ff$ and the GNN prediction for each of the cases, considered as the relative error $\varepsilon$, based on the $L2$-norm, defined as
\begin{equation}\label{eq:l2_norm}
\varepsilon = \dfrac{||\ff - \hat{\ff}||}{||\ff||} = \dfrac{\left [ \int_\Omega (\ff - \hat{\ff})^2 d\Omega \right ]^{\nicefrac{1}{2}}}{\left(\int_\Omega \ff^2  d\Omega\right)^{\nicefrac{1}{2}}}.
\end{equation}
The second metric employed arises from the necessity to assess the accuracy on the meanflow reconstruction $\hat{\vdir}$ from the GNN prediction with respect to the ground truth meanflow $\vdir$. It is defined as
\begin{equation}\label{eq:l2_norm_meanflow}
\delta = \dfrac{||\vdir - \hat{\vdir}||}{||\vdir||} = \dfrac{\left [ \int_\Omega (\vdir - \hat{\vdir})^2 d\Omega \right ]^{\nicefrac{1}{2}}}{\left(\int_\Omega \vdir^2  d\Omega\right)^{\nicefrac{1}{2}}}.
\end{equation}
In the previous equations $\Omega$ is the computational domain (Sec.~\ref{subsec:computational_domain}), $\hat{\ff}$ is the GNN prediction and $\ff$ is the closure term of the RANS equations coming from the DNS. Plugging $\hat{\ff}$ in Eq. \ref{Eq:RANS} and solving the inverse problem we can obtain $\hat{\vdir}$, a reconstruction of the meanflow based on the GNN prediction.

\subsection{Proof-of-concept training: flow past a cylinder flow}
\label{subsec:base_case}
Here, we consider the baseline results that in the following we will indicate as proof-of-concept (PC). The training dataset contains data obtained from DNS simulations of the flow past a cylinder; it is composed by $11$ pairs of meanflows $\mean \vdir$ (input) and Reynolds forcing stress $\ff$ (target), in the interval $50 \le Re \le 150$, with a stride of $\Delta Re = 10$. As previously mentioned, all the cases in the dataset exhibit a von Karman street instability in the wake of the bluff body.

\begin{figure}[t!]
    \centering
    \includegraphics[width=0.9\textwidth]{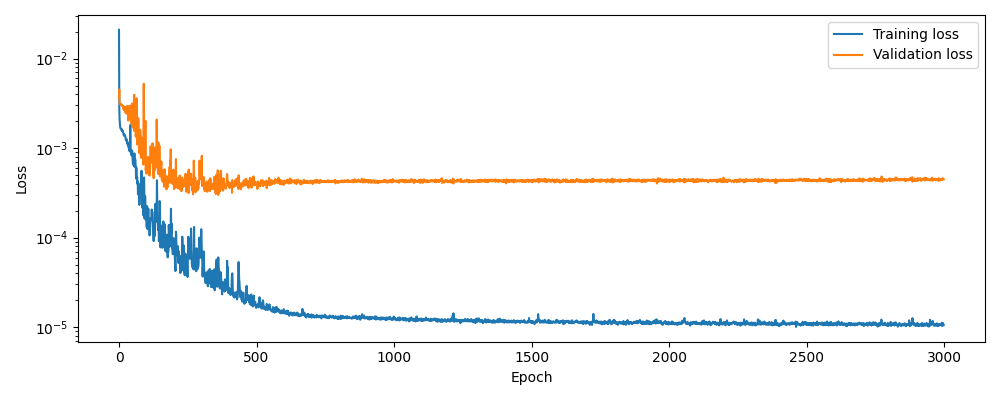}
    \caption{Curves for the training and validation loss for the proof-of-concept baseline, trained until $3000$ epochs. The training dataset is composed by $11$ meanflow-forcing pairs stacked at different Reynolds numbers in the range $50 \le Re \le 150$ with $\Delta Re = 10$. The validation set is formed by the test cases presented in Sec.~\ref{subsec:test_cases}.}
    \label{fig:cylinder_plot_loss}
\end{figure}

\smallskip
Fig.~\ref{fig:cylinder_plot_loss} shows the training and validation loss curves for this training approach. The training loss behaviour underlines the model's ability to capture the dynamics of the fluid flows and to learn the patterns in the training dataset. However, a significant gap can be observed between the training and validation loss curves. This discrepancy suggests a potential problem of overfitting to the training data. This problem is not unexpected, considering that the training dataset is composed exclusively of data obtained from simulations past the same geometry. Thus, when the GNN model is tested with unseen shapes or fluid dynamic conditions, the performance drops.  

\begin{figure}[t!]
    \centering
    \includegraphics[width=1\textwidth]{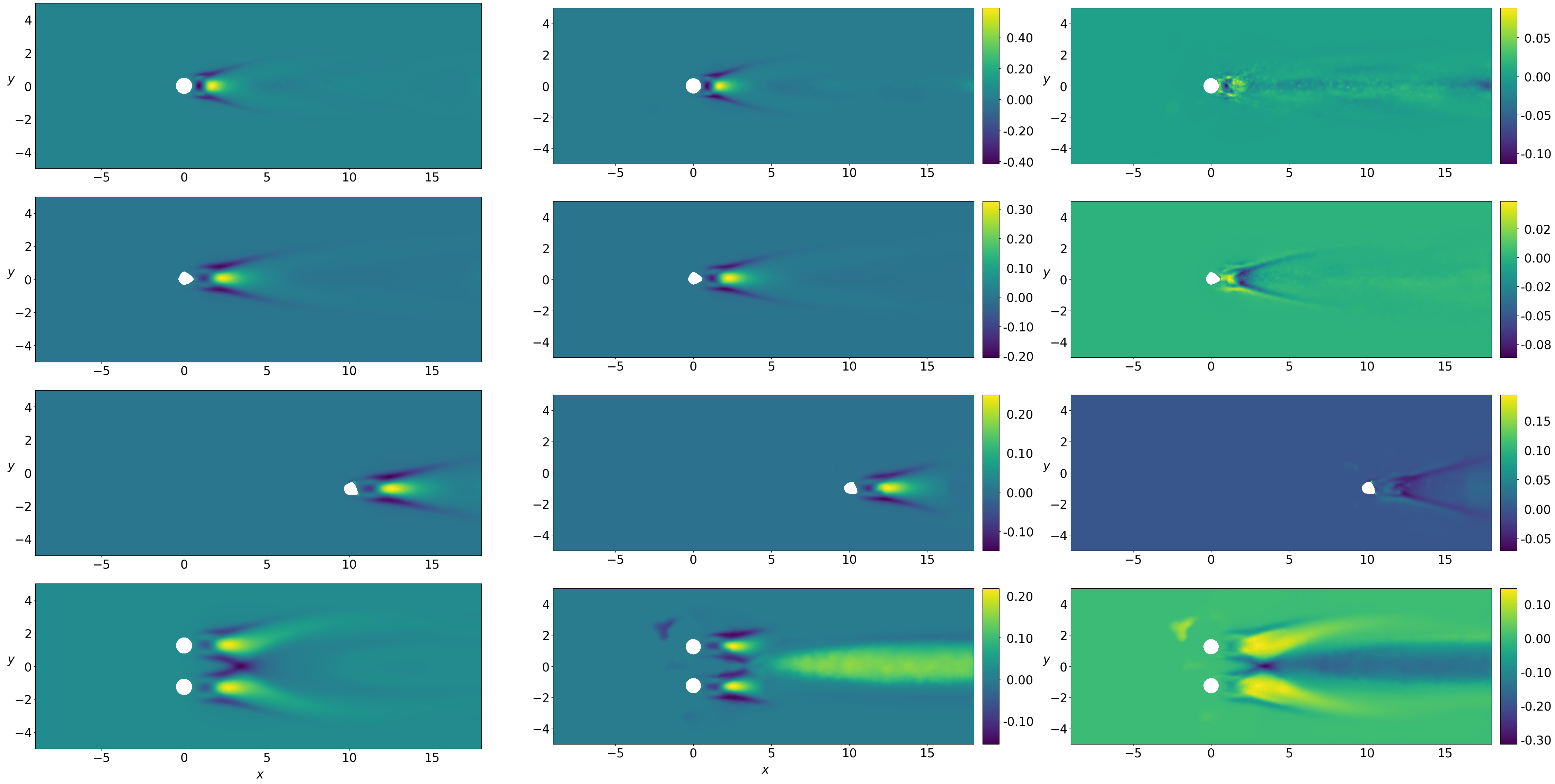}
    \begin{picture}(0,0)
        \put( -255, 230){$(a)$}
        \put( -255, 175){$(b)$} 
        \put( -255, 110){$(c)$} 
        \put( -255, 48){$(d)$} 
        \put( -165, 252){DNS}
        \put( -20, 252){GNN}
        \put( 130, 252){Difference}
    \end{picture}
    \caption{Stream-wise component comparison of the Reynolds stress tensor. The GNN's model is trained with a dataset composed by $11$ cases of cylinder shape bluff bodies, ranging in $50 \le Re \le 150$, $\Delta Re = 10$. $(a)$ {Case~1}, $Re=200$, cylinder bluff body shape; $(b)$ {Case~2}, $Re=120$, random shape bluff body; $(c)$ {Case~3}, $Re=100$, random shape shifted bluff body; $(d)$ {Case~4}, $Re=90$, flow past two side-by-side cylinders.}\label{fig:proof_of_concept_results}
\end{figure}

\smallskip
Regarding the test cases, in Case~1 (Fig.~\ref{fig:proof_of_concept_results}$a$), the GNN prediction shows good accuracy in reproducing the fluid structures at higher Reynolds number $Re=200$, with $\varepsilon=0.1153$. Fig.~\ref{fig:proof_of_concept_results}$(b)$ shows Case~2, where the difference between the GNN prediction and the DNS data is primarily visible in the near wake region. This region is crucial for the development of the unstable dynamics as it corresponds to the so-called wavemaker region \citep{giannetti2007structural}; for cases as such of non-standard geometrical shape, the flow vortex shedding behaviour can be quite different from the one observed in standard cylinder cases. The $\varepsilon$ norm in this case is $\varepsilon=0.2995$. Case~3 in Fig.~\ref{fig:proof_of_concept_results}$(c)$ tests the GNN generalization capabilities when the bluff body changes; the main discrepancies are again observed in the near wake of the bluff body. The $\varepsilon$ norm in this case is $\varepsilon=0.2084$. Finally, Case~4 in Fig.~\ref{fig:proof_of_concept_results}$(d)$ is particularly challenging as it tests generalization capabilities in presence of multiple bluff bodies. The GNN model's predictions capture the major features of the forcing stress of each of the cylinder. However, discrepancies are mainly observed in the area between the two cylinders and extending in the far wake region, where interactions between the Reynolds stress fields of the two cylinders occur. The $\varepsilon$ norm in this case is $\varepsilon=0.8704$, significantly higher compared to the previous cases. 

\smallskip
In all the cases discussed so far, the presence of some numerical errors in all the predictions is noted, but these are primarily attributed to statistical noise or inherent errors typical of neural network models. Neural networks indeed, by their nature, include elements of statistical uncertainty due to factors such as the stochastic nature of their training algorithms (e.g., random initialization of weights, batch selection during training), and the approximate nature of the model that represents the underlying physics.
Beside this aspect, this preliminary analysis based on the benchmark cases suggests that the GNN model performance could be further improved with a training set including a larger variety of bluff body shapes. Therefore, in what follows, we study the effects of the data augmentation by including in the training dataset random shaped bluff bodies with the aim of enhancing the model's robustness and generalization capabilities.

\subsection{Data augmentation and active learning: fluid flows past random geometries}
\label{subsec:random_shape}
In the following, we study the effects of data amount and quality on the training process and generalization properties of the GNN models. This study is structured into distinct steps.
\begin{enumerate}
\item \textbf{Data augmentation (DA)} (Sec.~\ref{subsubsec:data_augmentation}); this first approach involves dataset augmentation, meaning that we incorporate bluff bodies of random shapes  cases into the existing dataset, while maintaining the same dataset size used in the PC case, \textit{i.e.} $11$ cases. This method is aimed at diversifying the range of geometries used during the training of the model, without increasing the data volume.
\item \textbf{Dataset expansion} (Sec.~\ref{subsubsec:dataset_expansion}); in this second approach, we expand the dataset by including $33$ cases of bluff bodies of random shapes. We assess the quality of the used data in terms of sensitivity of the model to specific configurations by performing a $k$-fold validation.
\item \textbf{Active learning (AL) data selection} (Sec.~\ref{subsubsec:active_learning}); in this last approach, the training set is built by adding progressively data chosen by similarity criterion. The goal is to develop a surrogate model that can generalize at its best to unseen cases based on a given dataset.
\end{enumerate}

\subsubsection{Data augmentation with random shapes}
\label{subsubsec:data_augmentation}
In this phase of the study, we employ a {stratified random sampling} criterion to select shapes among the randomly generated configurations detailed in Sec.~\ref{subsec:random_shape_generation}. This latter involves the generation of the entire set of shapes for each Reynolds number in the interval $50 \le Re \le 150$, $\Delta Re = 10$ and randomly select $1$ shape for each value of the Reynolds number, for a total of $11$ shapes that compose the new training dataset. The chosen geometries are unique. On the selected shapes we perform DNS simulations in order to obtain the meanflow (input) and the forcing stress (target) used to train the GNN. 
\begin{figure}[t!]
    \centering
    \includegraphics[width=0.9\textwidth]{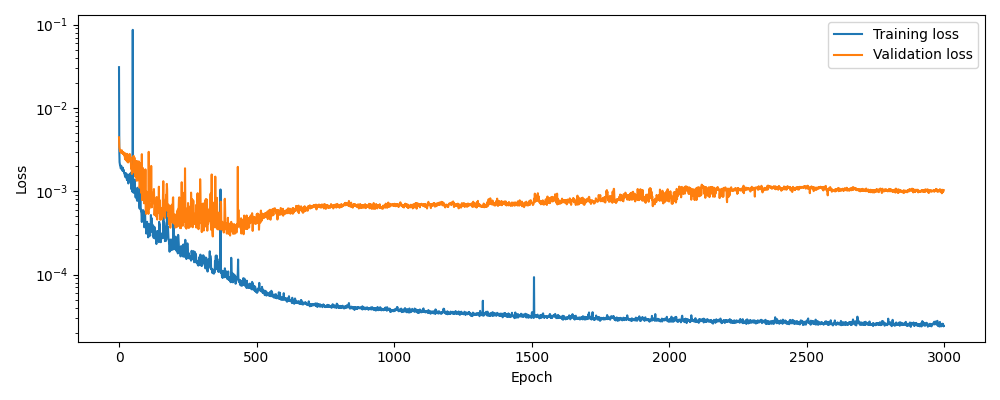}
    \caption{Curves for the training and validation loss for when DA is performed by introducing $11$ cases of flows past random geometries in the dataset. The Reynolds number varies in the range $50 \le Re \le 150$ with $\Delta Re = 10$. The training runs until $3000$ epochs. The validation set is formed by the test cases presented in Sec.~\ref{subsec:test_cases}.}\label{fig:random_shape_plot_loss}
\end{figure}

\smallskip
Fig.~\ref{fig:random_shape_plot_loss} shows the training and validation loss curves for this second training approach. The training loss demonstrates a consistent downward trend, highlighting the learning effectiveness of the GNN model. Notably, while there remains a significant gap between the training and validation loss curves, this gap is less pronounced when compared to the initial training approach (Fig.~\ref{fig:cylinder_plot_loss}). This reduced gap is indicative of diminished overfitting and suggests that the introduction of a wider variety of shapes and flow conditions into the training dataset leads to improved GNN model's generalization capabilities. 

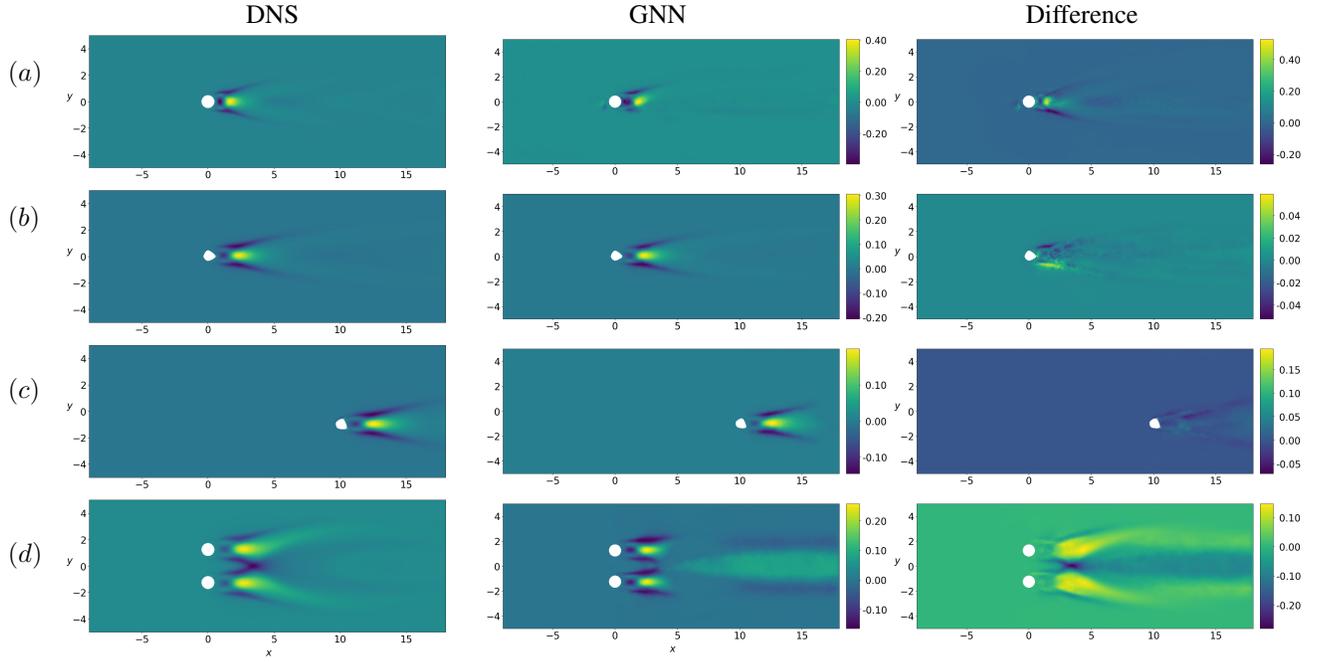
\begin{figure}[t!]
    \centering
    \includegraphics[width=1\textwidth]{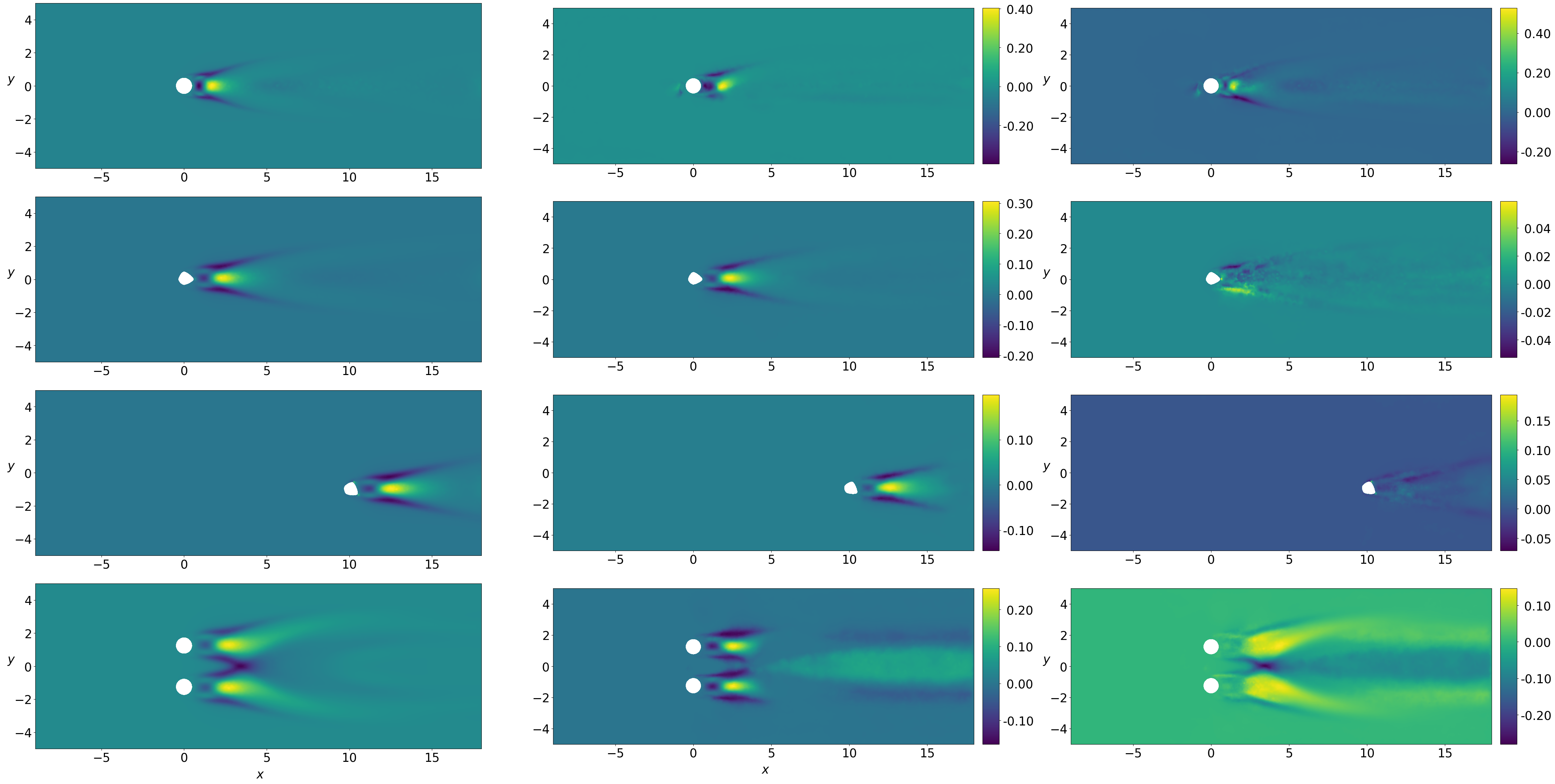}
    \begin{picture}(0,0)
        \put( -255, 230){$(a)$}
        \put( -255, 175){$(b)$} 
        \put( -255, 110){$(c)$} 
        \put( -255, 48){$(d)$} 
        \put( -165, 252){DNS}
        \put( -20, 252){GNN}
        \put( 130, 252){Difference}
    \end{picture}
    \caption{Stream-wise component comparison of the Reynolds stress tensor. The GNN's model is trained with a dataset composed by $11$ cases of random shaped bluff bodies, ranging in the interval $50 \le Re \le 150$, $\Delta Re = 10$. $(a)$ Case~1, $Re=200$, cylinder bluff body shape; $(b)$ Case~2, $Re=120$, random shape bluff body; $(c)$ Case~3, $Re=100$, random shape shifted bluff body; $(d)$ Case~4, $Re=90$, flow past two side-by-side cylinders.}\label{fig:random_shapes_results}
\end{figure}

\smallskip
Regarding the GNN predictions on test cases, Case~1 (Fig.~\ref{fig:random_shapes_results}$a$) shows larger discrepancies in the forcing prediction in the neighbourhood of the cylinder compared to the training approach discussed in Sec.~\ref{subsec:base_case}, $\varepsilon = 0.5033$. An interpretation for this result is given by the nature of the PC training dataset that solely relies on cylindrical geometries; thus, the resulting GNN model is specialized in the prediction of the cylinder flows. In contrast, the current training strategy involves flows past random shapes: from one hand, this choice broads the GNN's flexibility to capture diverse flow conditions by enhancing generalization capabilities; on the other hand, it simultaneously decreases the prediction accuracy for cylinder shaped cases. 

Case~2 (Fig.~\ref{fig:random_shapes_results}$b$) and Case~3 (Fig.~\ref{fig:random_shapes_results}$c$) show mainly numerical and statistical noise in the difference fields while the main features of the Reynolds stress tensor are well reproduced in the prediction. Contrary to what is observed in Case~1, this training approach leads to an efficient GNN model in predicting the features of the flows past random shaped bluff bodies. Error norm is comparable for the two test cases, namely $\varepsilon = 0.2063$ for Case~2 and $\varepsilon = 0.1999$ for Case~3. Finally, Case~4 is characterized by inaccuracies primarily located in the wake region similar to the one already observed in the PC case (for a reference, the reader can compare Fig.~\ref{fig:proof_of_concept_results}$d$ and Fig.~\ref{fig:random_shapes_results}$d$). However, it's important to note that these errors are quantitatively less significant to those observed in the first training approach, with $\varepsilon = 0.6312$.

\subsubsection{Dataset Expansion}
\label{subsubsec:dataset_expansion}
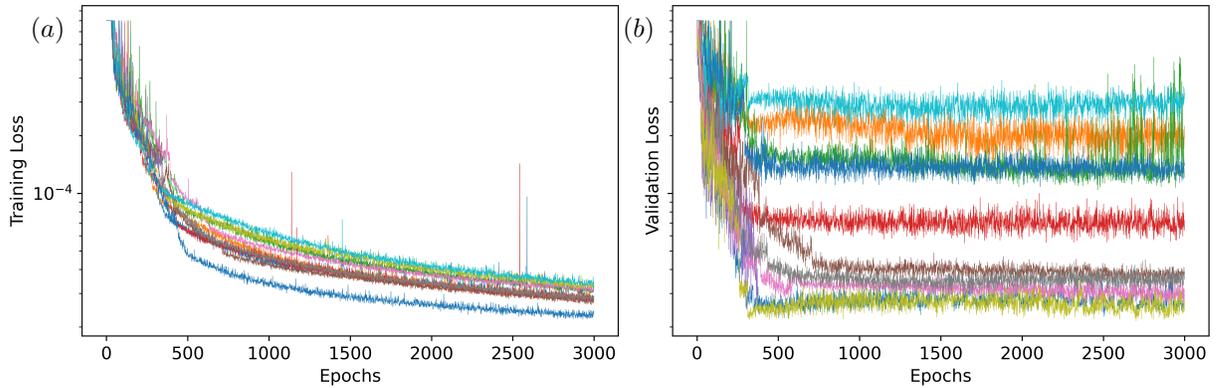
\begin{figure}[t!]
    \centering
    \includegraphics[width=\textwidth]{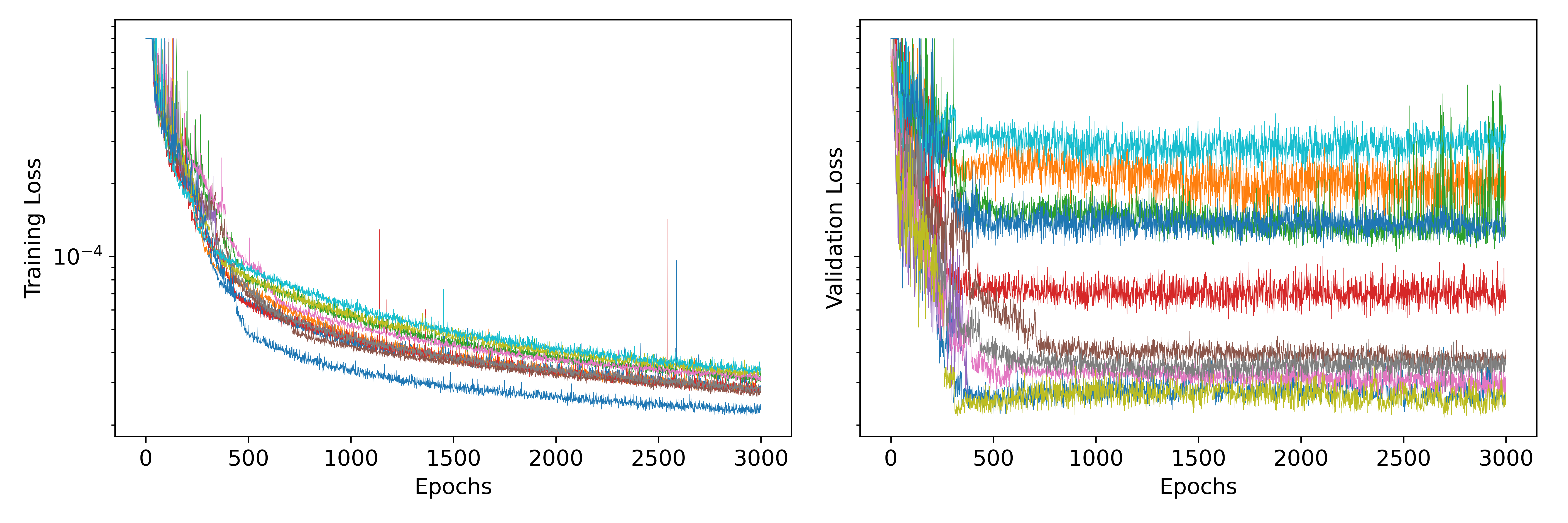}
    \begin{picture}(0,0)
        \put(-220, 150){$(a)$}
        \put( 4, 150){$(b)$}
    \end{picture}
    \caption{Training and validation loss curves for models trained using $10$ $k$-fold up to $3000$ epochs. $(a)$ shows the training curves, $(b)$ the corresponding validation curves for each fold.}\label{fig:train_val_curve}
\end{figure}
A common technique to improve generalization in neural network models is to enlarge the volume of data the model is trained with. This approach reduces the risk for the NN model to overfit to specific conditions observed in a small dataset and is more likely to capture the underlying phenomena. Therefore, we study the effect of tripling the amount of data in the training dataset. With the same {stratified random sampling} approach described in Sec.~\ref{subsubsec:data_augmentation}, we select $3$ cases for each of the $11$ Reynolds numbers chosen in the interval of reference, resulting in a $33$ cases training dataset. In order to analyse the sensitivity of the GNN model's generalization capabilities with respect to the training dataset, we employ a {$k$-fold} validation test. First, we divide the training dataset into $k$ groups and train $k$ different models. For each of the $k$ models, we use $k-1$ groups as training dataset and the $k$-th remaining group as the validation dataset. Although this approach is not feasible in practical applications, it is very informative for assessing preliminarily the impact of the quality of data on the final prediction.

\smallskip
Fig.~\ref{fig:train_val_curve}$(a)$ shows the training curves of the $10$ models, one for each of the $k$ folds, while Fig.~\ref{fig:train_val_curve}$(b)$ displays their corresponding validation curves. The training curves are rather close, independently from the chosen dataset, thus suggesting a rather consistent behaviour of the chosen GNN architecture. However, the validation loss varies significantly. This variability can be attributed to the distribution of the training dataset. Inconsistent or unbalanced data distributions may lead the model to being trained on subsets of data that do not adequately represent the overall dataset, thus impacting negatively on the validation results. To address this challenge and identify the most effective dataset, we introduce AL in the following.

\subsubsection{Active Learning data selection}
\label{subsubsec:active_learning}
AL allows to dynamically adjust the dataset, targeting the data that contribute most significantly to the model's generalization capability. We exploit a {similarity criterion} from the GNN perspective, following the study by \cite{input_similarity}. The key idea is to avoid including in the training dataset cases that do not lead to any significative improvement of the model from the generalization viewpoint; instead, in order to induce diversity in the GNN model, we need to select cases that are as most as different from each other from the GNN {perspective}.
We can achieve this goal by comparing for each of the available data pairs the vector gradient of the cost function with respect to the learnable parameters of the GNN $\mathbf{\theta}$ by means of scalar products. As comparison metric, we employ the {cosine similarity} angular distance
\begin{equation}\label{eq:cosine_similarity}
    \cos (\beta ) =   \dfrac {\mathbf{a} \cdot \mathbf{b}} {\left\| \mathbf{a}\right\| \left\| \mathbf{b}\right\|},
\end{equation}
where $\mathbf{a}\in \mathbb{R}^m$ and $\mathbf{b}\in \mathbb{R}^m$ are two generic $m$-dimensional vectors and $\beta$ is the angle between them. The basic idea is that each gradient vector steers the neural network's state to a specific direction within the solution space. Therefore, including multiple data points whose gradients point in the same direction might be redundant. Instead, our aim is to select and include in the training dataset as many different directions as possible, in order to explore more extensively the solution space (see also \ref{appC:similarity_criteria}).  

\smallskip
The selection process stops when a predefined {similarity} threshold is reached, such that only the pairs exhibiting a similarity below this threshold are added in the dataset. Conversely, cases that show a similarity value exceeding the threshold are discarded. However, the {similarity} threshold does not have a direct, interpretable meaning and needs to be tuned as an input parameter of the process, according to the required performances. In general, a lower similarity threshold means that more cases will be included in the training dataset. While this can enhance models behaviour to diverse scenarios, it may also lead to a more complex and time-consuming training process. On the other hand, a too restrictive similarity threshold might exclude potentially valuable training data. Therefore, an optimal similarity threshold is chosen as a trade-off between the training computational costs and the model generalization performances.
\begin{figure}[t!]
    \centering
    \includegraphics[width=0.9\textwidth]{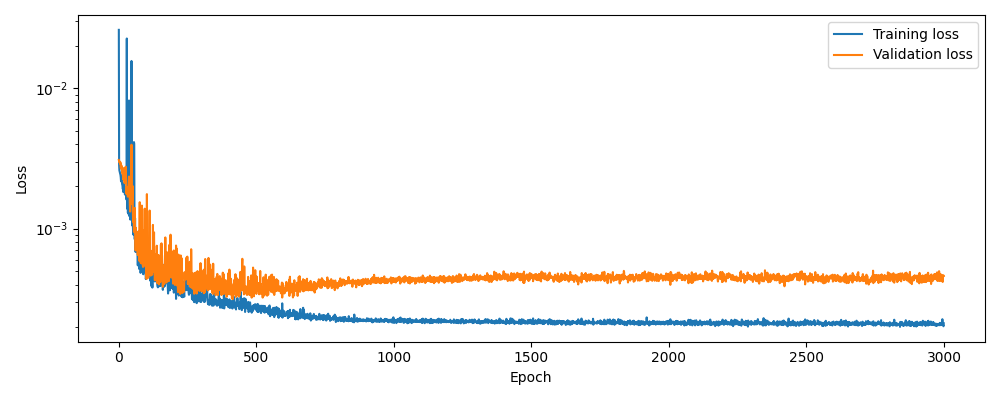}
    \caption{Training and validation loss curves are shown when AL with similarity threshold $0.8$ is performed until $3000$ epochs. The training dataset is formed by $6$ random shaped bluff bodies. Their Reynolds number varies in the interval $50 \le Re \le 150$. The validation set is formed by the test cases presented in Sec.~\ref{subsec:test_cases}.} \label{fig:input_sim_plot_loss}
\end{figure}

\begin{figure}[t!]
    \centering
    \includegraphics[width=1\textwidth]{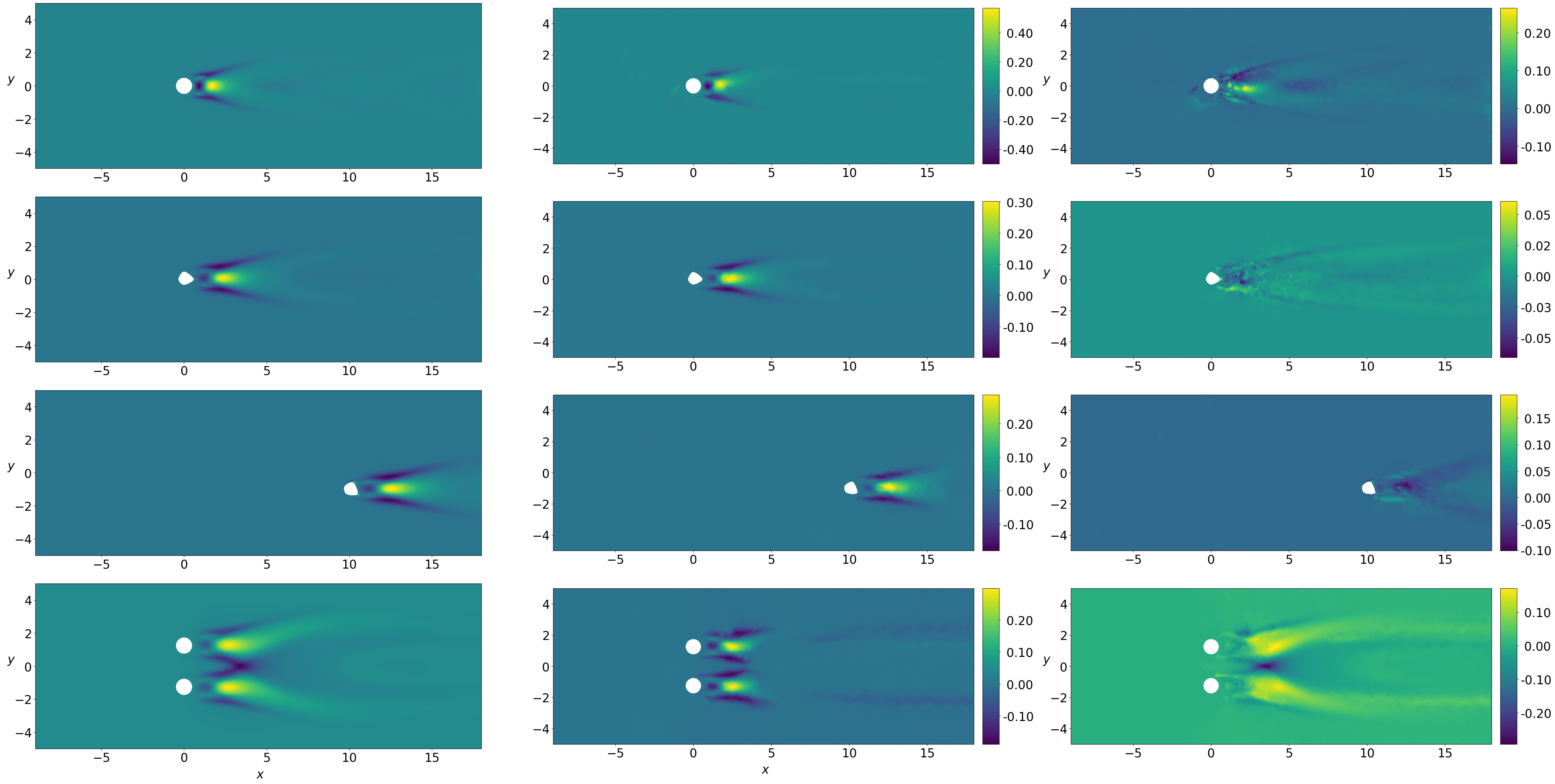}
    \begin{picture}(0,0)
        \put( -255, 230){$(a)$}
        \put( -255, 175){$(b)$} 
        \put( -255, 110){$(c)$} 
        \put( -255, 48){$(d)$} 
        \put( -165, 252){DNS}
        \put( -20, 252){GNN}
        \put( 130, 252){Difference}
    \end{picture}
    \caption{Stream-wise component comparison of the Reynolds stress tensor. The GNN's model is trained with a dataset composed by $6$ cases of bluff bodies of random shape selected with the AL approach, ranging in the interval $50 \le Re \le 150$, $\Delta Re = 10$. $(a)$ Case~1, flow past a cylinder at $Re=200$; $(b)$ Case~2, flow past a random shaped bluff body at $Re=120$; $(c)$ Case~3, random shaped bluff body at $Re=100$, shifted downstream; $(d)$ Case~4, flow past two side-by-side cylinders at $Re=90$.} \label{fig:input_sim_results}
\end{figure}
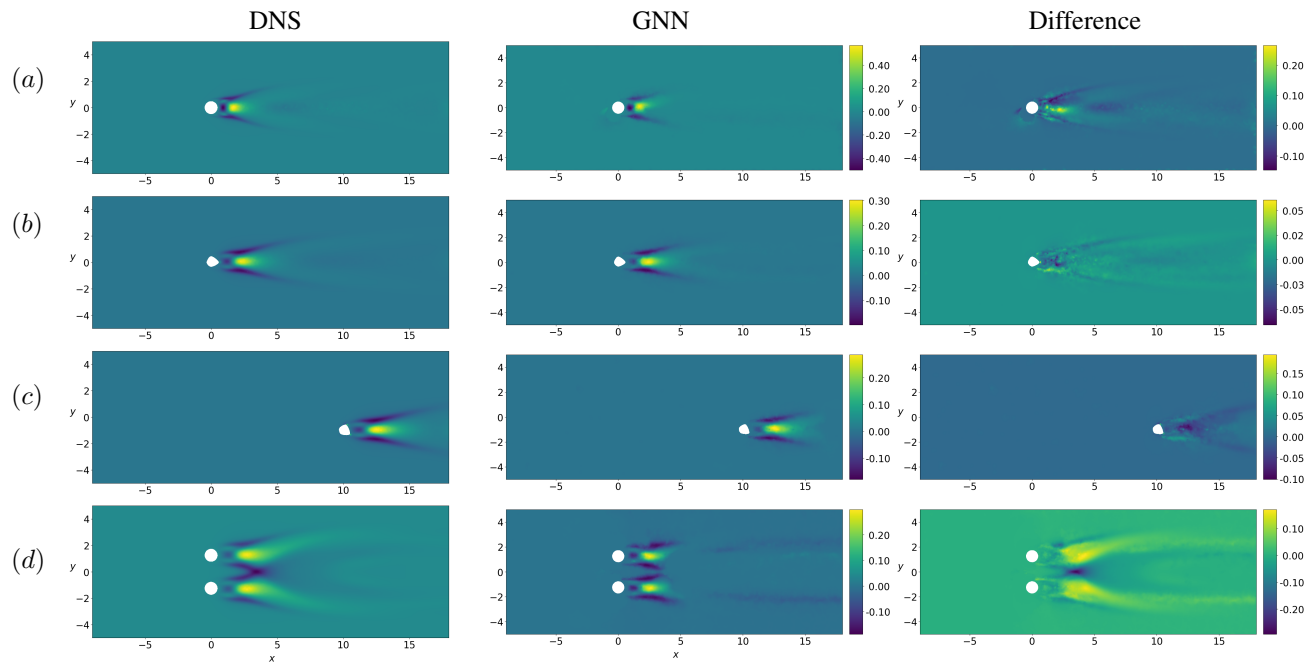

\smallskip
In our study, we explore three different similarity threshold values, $\cos (\beta ) \in [0.7, 0.8, 0.9]$. This approach allows to observe the impact of varying levels of data inclusion on the model's training and performance. Results shown in Fig.~\ref{fig:input_sim_results} and  Fig.~\ref{fig:input_sim_plot_loss} are obtained with a similarity threshold of $0.8$, which results in only $6$ selected random bluff body cases for the training dataset; the results for the other threshold values are detailed in Sec.~\ref{sec:comparison}). Fig.~\ref{fig:input_sim_plot_loss} reports the training and validation loss curves: the reduced disparity observed between the two curves -- compared to the results from the previous learning approaches -- underscores a notable decrease in the issue of overfitting. The training is initiated with a single case in the training dataset, specifically the cylinder case at $Re = 120$. 

Once the convergence of the vector gradients is reached (see \ref{appC:similarity_criteria}), a new case in training set is selected; since this process is a seek-and-include algorithm that enlarges the dataset at each step, two different approaches can be applied. The first strategy is inspired by the {curriculum learning} framework \citep{bengio2009curriculum}, where each selected case is added to the ongoing training. Thus, the model is progressively updated. 
A second approach, applied in this work, consists in reinitializing the GNN weights every time the dataset is enlarged with a new pair. The rationale behind this choice is based on the dynamics of the solution space, that changes when new data are added in the training set; in this sense, there is no guarantee that the new minima will be "closer" to the previous ones than to the initialization point of the GNN. Nonetheless, we observed that for the analysed flow cases the two training strategies lead to negligible differences in the final results.

\smallskip
Considering the benchmarks for the model assessment, Case~1 leads to $\varepsilon = 0.5507$ between the GNN predictions of the Reynolds stress tensor and the DNS ones. The error is $\varepsilon = 0.1518$ for the random shape bluff body case (Case~2), $\varepsilon = 0.3595$ for the downstream shifted random shape bluff body (Case~3) and $\varepsilon = 0.5243$ for the two side-to-side cylinder configuration (Case~4). The results respect the simmetry of the solutions with respect of the $y$ axis and only minor inconsistencies can be observed in the far wake; we note that the errors are concentrated in the region immediately downstream of the bluff body, but we interpret these errors as numerical or stochastic in nature rather than being prediction inaccuracies of the fluid structures.

\subsection{Quantitative comparison}
\label{sec:comparison}
In conclusion, we summarize the results in Tab.~\ref{tab:results_comparison}. A direct comparison between the first two training approaches (PC and DA) reveals that enriching the training dataset with random geometries enhances the generalization capabilities of the GNN. This behaviour is not unexpected, although an exception is observed when extrapolating at higher Reynolds numbers (Case~1), where we observe that the PC approach outperforms DA. As already discussed, this behaviour can be understood by observing that the GNN model trained following the approach PC is specialized in predicting the flow past a cylinder as it is trained on this specific geometry.\\ On the other hand, the AL training approach allows to obtain GNN models demonstrating overall superior performance in terms of generalization capabilities as compared to PC and DA approaches, in particular for Case~2 and Case~4, while in Case~3 performance are essentially comparable. The mean error decreases in all the $4$ test cases in the AL approach, indicating an improvement in the global performance of the model. An important aspect regarding the active selection of data for the training is the amount of data. Good performances are achieved with only $6$ pairs when a threshold $0.8$, although it is also possible to observe a strong variation in the total number of pairs used for the training as a function of the chosen threshold. It is crucial to observe, however, that for all the AL cases the introduction of a selection criterion guarantees that the chosen data points lead to good performance in terms of generalization.\\
When comparing our work with existing literature, the notable aspect is that our approach achieves comparable accuracy with significantly fewer training cases. For instance, in \citep{chen2021graph} the authors utilize $2000$ cases in their training dataset for steady-state incompressible flow around a cylinder at $Re = 10$. In contrast, we use only 19 cases for the largest dataset used and still manage to generalize to different Reynolds numbers and bluff body positions, with comparable accuracy results. In \citep{Lee_2019}, $500k$ cases are used for the training dataset although an unsteady flow is predicted using CNN. It is worth noting that the use of a GNN architecture enables to generalize on different geometries and $Re$ numbers, an aspect that is not addressed in \citep{Lee_2019}. Finally in \citep{Thuerey_2020}, a GNN is employed for predicting time--averaged steady flow. The dataset consists of $12800$ data points while the GNN model has a complexity of over $30M$ parameters, prohibitive for most practical applications. Our GNN architecture, on the contrary, can count up to approximately $900k$ parameters. In conclusion, we believe that the combination of GNN models and active learning makes our method more accessible and practical for broader applications thanks to the parsimony of the data requirements.
\begin{table}[t!]
\resizebox{1\textwidth}{!}{
\begin{tabular}{|c|cccccccccc|}
\hline
\multirow{3}{*}{\textbf{Cases}} &\multicolumn{2}{c|}{\textbf{PC}} &\multicolumn{2}{c|}{\textbf{DA}} &\multicolumn{6}{c|}{\textbf{AL}}
\\ 
&\multicolumn{2}{c|}{}  &\multicolumn{2}{c|}{}  
&\multicolumn{2}{c|}{\textbf{0.7}}  &\multicolumn{2}{c|}{\textbf{0.8}} &\multicolumn{2}{c|}{\textbf{0.9}}
\\ \hline

&\multicolumn{1}{c|}{\textit{$\varepsilon$}}             
&\multicolumn{1}{c|}{\textit{$\delta$}}
&\multicolumn{1}{c|}{\textit{$\varepsilon$}}             
&\multicolumn{1}{c|}{\textit{$\delta$}}
&\multicolumn{1}{c|}{\textit{$\varepsilon$}}             
&\multicolumn{1}{c|}{\textit{$\delta$}}
&\multicolumn{1}{c|}{\textit{$\varepsilon$}}             
&\multicolumn{1}{c|}{\textit{$\delta$}}
&\multicolumn{1}{c|}{\textit{$\varepsilon$}}             
&\multicolumn{1}{c|}{\textit{$\delta$}}
\\ \hline
\textit{Case~1}     
&\multicolumn{1}{c|}{\textbf{0.1153}} %epsilon
&\multicolumn{1}{c|}{\textit{0.0118}} %delta
&\multicolumn{1}{c|}{0.5033} %epsilon 
&\multicolumn{1}{c|}{0.1987} %delta
&\multicolumn{1}{c|}{0.8671} %epsilon
&\multicolumn{1}{c|}{0.1859} %delta
&\multicolumn{1}{c|}{0.5507} %epsilon        
&\multicolumn{1}{c|}{0.1165} %delta
&\multicolumn{1}{c|}{0.4132} %epsilon     
&\multicolumn{1}{c|}{0.1162} %delta
\\ \hline
\textit{Case~2}     
&\multicolumn{1}{c|}{0.2995} 
&\multicolumn{1}{c|}{0.0280} %delta
&\multicolumn{1}{c|}{0.2063}
&\multicolumn{1}{c|}{0.0160} %delta
&\multicolumn{1}{c|}{0.1256} 
&\multicolumn{1}{c|}{\textit{0.0079}} %delta
&\multicolumn{1}{c|}{0.1518}
&\multicolumn{1}{c|}{0.0166} %delta
&\multicolumn{1}{c|}{\textbf{0.1124}}
&\multicolumn{1}{c|}{0.0092} %delta
\\ \hline
\textit{Case~3}     
&\multicolumn{1}{c|}{0.2084} 
&\multicolumn{1}{c|}{0.0131} %delta
&\multicolumn{1}{c|}{\textbf{0.1999}}
&\multicolumn{1}{c|}{0.0090} %delta
&\multicolumn{1}{c|}{0.3058} 
&\multicolumn{1}{c|}{0.0107} %delta
&\multicolumn{1}{c|}{0.3595}
&\multicolumn{1}{c|}{0.0183} %delta
&\multicolumn{1}{c|}{0.2123}       
&\multicolumn{1}{c|}{\textit{0.0082}} %delta
\\ \hline
\textit{Case~4}     
&\multicolumn{1}{c|}{0.8707} 
&\multicolumn{1}{c|}{0.2683} %delta
&\multicolumn{1}{c|}{0.6312}    
&\multicolumn{1}{c|}{0.1701} %delta
&\multicolumn{1}{c|}{0.6751} 
&\multicolumn{1}{c|}{0.1930} %delta
&\multicolumn{1}{c|}{\textbf{0.5243}}     
&\multicolumn{1}{c|}{\textit{0.1565}} %delta
&\multicolumn{1}{c|}{0.5719}       
&\multicolumn{1}{c|}{0.1653} %delta
\\ \hline \hline
\textit{Training data}     
&\multicolumn{2}{c|}{11} 
&\multicolumn{2}{c|}{11}    
&\multicolumn{2}{c|}{3} 
&\multicolumn{2}{c|}{6}     
&\multicolumn{2}{c|}{19}           
\\ \hline
\end{tabular}
}
\caption{Comparison on the $4$ cases defined as benchmark for the three different training approaches: the proof-of-concept (PC) training (Sec.~\ref{subsec:base_case}), the data augmentation (DA) training (Sec.~\ref{subsubsec:data_augmentation}), and the active learning (AL) strategy (Sec.~\ref{subsubsec:active_learning}). For the latter, we consider three similarity threshold values $\cos (\beta) \in [0.7, 0.8, 0.9]$. The chosen metric $\varepsilon$ and $\delta$ are defined respectively in Eq. \ref{eq:l2_norm} and Eq. \ref{eq:l2_norm_meanflow}. In the last row the number of pairs used during the training process is reported.}
\label{tab:results_comparison}

\end{table}

\section{Conclusion}
\label{sec:conclusions}
The application of machine learning techniques in fluid mechanics is often characterized by shortcomings such as overfitting, lack of robustness of the prediction with respect to unseen cases and data-hungriness. In this article, we proposed a novel approach that combines a neural architecture based on Graph Neural Networks (GNN), numerical solvers based on Finite Element Method (FEM) and an active learning procedure in order to tackle some of these limitations. Here, we consider a data-assimilation schemes that does not rely on an optimization process and use as baseline equations the Reynolds-averaged Navier-Stokes (RANS) equations. GNN models are trained as a surrogate to predict the forcing/closure term, obtained as an output of the supervised learning, while a given mean flow serves as input. The GNN architecture is particularly suitable in this study due to its adaptability to unstructured meshes and its generalization capability, as compared to other literature approaches. Moreover, this architecture allows frugal training within the low-data limit, as compare to alternative, more expensive in terms of required data, architectures. 

\smallskip
A two-fold interface between FEM and GNN environment has been developed to transform a FEM vector field into a numerical tensor that can be handled by a NN structure and vice versa, preserving critical information throughout the process. As test-bed, we focussed on two dimensional, incompressible flows past obstacles at low Reynolds numbers, namely in the range $50 \le Re \le 150$. At these regimes, the presence of obstacles triggers instabilities developing in unsteady flows. We started by studying a cylindrical geometry in the range $50 \le Re \le 150$ as initial benchmark, in order to assess the extent to which a model based on this training dataset can be used also for unseen cases. Not surprisingly, we found good performance at unseen Reynolds number for the cylinder case. On the other hand, when the flow around the bluff bodies of random geometry is considered, it is observed lack of accuracy in the prediction and overfitting. 

In order to tackle these limitations, we explored the impact of the training data on the generalization capabilities of the GNN in terms of quantity and quality of data. First, we considered an extended dataset. Our results indicate that the quality and volume of data notably affect the spectrum of unseen cases on which the model can generalize to. Particularly, the inclusion of diverse fluid flow conditions into the training dataset improves the overall generalization capabilities for the vast majority of cases. Finally, we introduced an active learning data selection criterion based on the analysis of gradient similarity, with the aim of building a dataset extending the distribution of the data. At the best of authors knowledge, this is one of the first applications in the community of fluid mechanics where a systematic selection of the data is performed addressing the generalization of the NN model prediction to unseen cases by maximizing the quality of the predictions and at same time minimizing the amount of data used in the training set. The results clearly indicate the possibility of improving the performance of the model, also in terms of generalization, while keeping a rather small amount of data in the training set. The criterion is especially relevant in the contexts where computational resources for training surrogate models are limited, and a trade off between accuracy of the predictions and training computational cost is sought. It is stressed that the datasets used during the training process are relatively small as compared to other approaches appeared in ML literature, as the most expensive one consists of less than 20 pairs of snapshots. These datasets are selected through a criterion that minimizes the number of data points required and is robust enough to be applied to larger datasets to efficiently reduce them while maintaining essential information.

\smallskip
We are currently extending our findings by incorporating physical constraint in the learning loop through the adjoint equations associated with the assimilation loop. This integration aims to further refine the predictive performance and generalization capabilities of the GNN leveraging the physics constraints. The final goal of these data-assimilation schemes is to adapt our training approach to cases where solely limited or corrupted measurements of the flow are available, such as those based on sparse probe measurements, noisy or incomplete data.

\paragraph{Acknowledgements} The PhD fellowship of M.~Quattromini is supported by the Italian Ministry of University. This study has been partially funded under the National Recovery and Resilience Plan (NRRP), Mission 4 Component 2 Investment 1.3 - Call for tender No. 1561 of 11.10.2022 Project code PE0000021, Project title {Network 4 Energy Sustainable Transition – NEST}, funded by the European Union – NextGenerationEU. The support from the {Agence Nationale de la Recherche} (ANR) through grant ANR-21-REASON is gratefully acknowledged.             

\begin{appendices}
\section{Numerical simulations details}
\label{appA:numerical_sim_details}
Time-resolved numerical simulations were performed using a \verb}python} code based on the \verb}FEniCS}~\citep{alnaes2015fenics} library. The inputs and the outputs of the Graph Neural Network model were obtained by averaging these data on-the-fly.  More details on the numerical scheme can be found in \cite{FEM_book}; the main script is based on the code described in \cite{guegan2022control}, where an extensive discussion is reported. Here we summarize the essential aspects and provides details on the validation.
\smallskip
From the numerical viewpoint, the spatial discretization is obtained by introducing the weak formulation based on the finite element method (FEM). In particular, the finite element used is the Taylor-Hood element, with second order elements P2 for velocity and first order elements P1 for pressure. The implemented spatial integration scheme reads as
\begin{eqnarray}
\begin{cases}
\begin{array}{r l}
\left(\dfrac{3\vdir^n-4\vdir^{n-1} +\vdir^{n-2}}{2\Delta t}\right)
+(\vdir^{n-1}\cdot\nabla)\vdir^n+(\vdir^n\cdot\nabla)\vdir^{n-1} & \\
-(\vdir^{n-1}\cdot\nabla)\vdir^{n-1}-\dfrac{1}{Re}\Delta\vdir^{n}+\nabla p^{n}&= \bf{0} \\
\nabla\cdot\vdir^{n} &=0,
\end{array}
\end{cases}
\end{eqnarray}
where $\vdir=(u,v)^T$ represents the velocity vector, $p$ the pressure and $Re$ the Reynolds number. The time marching is performed by second order backward differentiation formula (BFD): the $n$ apex indicates the values of a quantity at the current time, with $n-1$ the values at the previous time step and $n-2$ its values two time steps before. The time step -- denoted as $\Delta t$ -- it is chosen for granting the Courant-Friedrichs-Lewy condition, $CFL \le 0.5$ for each of the random generated shapes. 
\begin{figure}[t]
\includegraphics[width=0.7\textwidth]{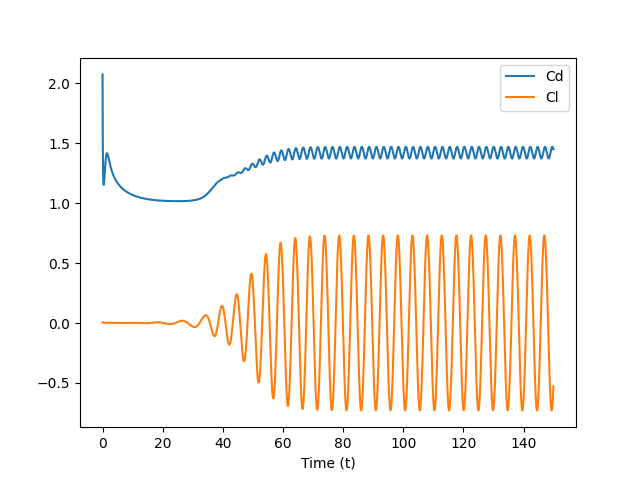}
\caption{Drag coefficient $C_d$ and lift coefficient $C_l$ of flow past a cylinder.}
\label{fig:drag_lift}
\end{figure}
The convective term is treated using the Newton and Picard methods, such that a linear system of equations is solved in each step of the temporal iteration.
The basic mesh is based on the one used for the cylinder flow, our reference case; the mesh is refined in the most sensitive regions around the obstacle and in the wake past the cylinder \citep{giannetti2007structural}: these are the areas where high refinement is required for increasing the accuracy of the computations in terms of correct frequencies and growth rate of the instabilities. Beside the well known dynamics, the rich literature allows a detailed comparison for validation purposes. Thus, numerical solver and meshes were validated using the cylinder flow, in particular by comparing the drag coefficient $C_d$ and the lift coefficient $C_l$. Fig.~\ref{fig:drag_lift} shows the evolution in time of $C_d$ and $C_l$, while the table Tab.~\ref{tab:drag_lift_comparison} shows the numerical comparison of (time-averaged) $C_d$ and $C_l$ (amplitude) with respect to values in literature. The first coefficient compares remarkably well with the values in literature, while a slightly higher $C_l$ is found: this can be related to the chosen numerical box that is slightly small along the $y$ direction, thus producing some numerical blockage effects; nonetheless, for the nature of the present work, the results are satisfactory. Finally, the chosen bluff bodies are characterized by mechanisms resembling the phenomena observed in the cylinder flow, thus characterized by comparable time and space scales; the corresponding meshes are characterized by a refinement similar to the one adopted for the reference case.
\begin{table}[t!]
\centering
\begin{tabular}{ccc} \hline
Cylinder & $C_d$ average & $C_l$ amplitude \\ \hline
Etienne and Pelletier & 1.36 & 0.67 \\
Li et al. & 1.34 & 0.69 \\
Liu et al. & 1.31 & 0.69 \\
Present result & 1.32 & 0.73 \\ \hline
\end{tabular}
\caption{Comparison of $C_d$ average and $C_l$ amplitude}%
\label{tab:drag_lift_comparison}
\end{table}
\section{Hyperparameters optimization}
\label{appB:GNN_optimization}
Numerous parameters define the structure of a neural network, usually denoted with the term {hyperparameters}. We can distinguish between {model} hyperparameters and {process} hyperparameters.
\begin{itemize}
    \item A model hyperparameter defines the {capacity} of the neural network, \emph{i.e.} the ability of the model to represent functions of high complexity. Thus, the capacity is directly related to the possibility of approximating a large variety of nonlinear functions.
    \item A process hyperparameter defines the training phase. Tuning these hyperparameters deeply modify the duration of the training, its computational costs and the way the weights are adjusted while the model evolves.
\end{itemize}
These terms are defined \textit{a-priori}, before training the model, thus they need to be manually tuned as they can't be adjusted or learnt during the training process. In general, this does not pose problems in deep learning when the expressivity of the NN (\textit{e.g.} the number of neurons, the number of layers, etc.) is enough to represent the complexity of the problem under investigation. On the other hand, since the GNN is trained to fulfil a specific task and because the computational cost of the GNN inference has to be affordable compared to the most sophisticated turbulence models available today, we tried to keep the GNN as parsimonious as possible. 

To this end, the hyperparameters defining the architecture require optimization. Standard gradient based optimizers cannot be employed when dealing with integer numbers, like the number of neurons or layers. For this purpose gradient-free algorithms can be used. There exists dedicated libraries that can automate the tuning process through all the possible sets of hyperparameters by trying and appropriately pruning the unpromising sets of them. In this work we apply the library \verb{Optuna{ \citep{optuna_2019}, an open-source package that combines efficiently searching and pruning algorithms. By exploring the complex solution hyperspace, a number of combinations of hyperparameters is found, among which the one outperforming the others in terms of monitored validation metrics is given by the following set
\begin{enumerate}
\item Embedded dimension, $35$
\item Number of GNN layers, $k = 40$
\item Update relaxation weight, $\alpha = 6\times 10^{-1}$
\item Loss function weight, $\gamma = 0.1$
\item Learning rate, $LR = 3\times 10^{-3}$, as maximal/starting value.
\end{enumerate}
\section{Similarity criteria algorithm details}
\label{appC:similarity_criteria}
The {similarity criteria} algorithm designed to compare different data from the neural network perspective is based on the analysis of the vector gradients of a metric function (Eq.~\ref{eq_metric_function}) with respect to the $\mathbf{\theta}$ parameters of the neural network. In particular, the similarity comparison between two generic $m$-dimensional vectors $\textbf{a} \in \mathbb{R}^m$ and $\textbf{b} \in \mathbb{R}^m$ is computed using the {cosine similarity}, defined as
\begin{equation}\label{eq.cosine_similarity}
\cos (\beta ) = \dfrac {\mathbf{a} \cdot \mathbf{b}} {\left\| \mathbf{a}\right\| \left\| \mathbf{b}\right\| },
\end{equation}
where $\beta$ is the angle between the two vectors. In this context the metric we use is the Mean Absolute Error (MAE), a piecewise linear function defined as
\begin{equation}\label{eq_metric_function}
MAE = \sum_{i=1}^{n_i}|x_i-y_i|,
\end{equation}
in which $x_i$ is the NN prediction on the node $i$, $y_i$ the ground truth and $n_i$ the number of nodes.

\smallskip
\begin{figure}[t]
\centering
\includegraphics[width=1\textwidth]{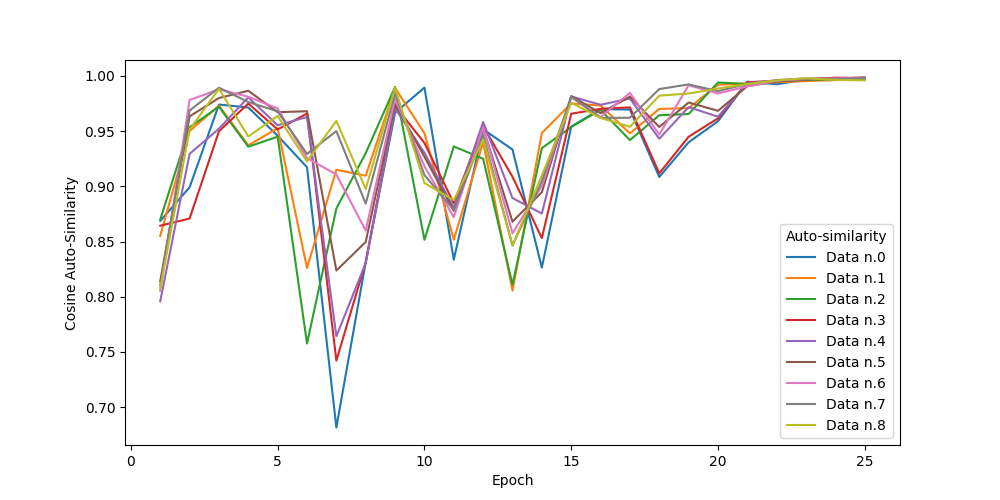}
\caption{Visualization of gradient auto-similarity convergence over multiple training epochs using MAE for 9 distinct cases within the training data set.}\label{fig:autosimilarity}
\end{figure}
The choice of MAE is crucial for our analysis. In scenarios where the Mean Squared Error (MSE) is used, we empirically observe marked oscillations in the direction of the auto-similarity of the vector gradient throughout successive training epochs. Conversely, when employing MAE, the auto-similarity of the vector gradient tends to approach unity, suggesting a stable direction in the solution space, for the data under analysis . A visual representation of the consistent convergence of the vector gradient auto-similarity as function of the epochs is shown in Fig.~\ref{fig:autosimilarity}, for a training process involving $9$ cases in the training dataset. Training begins with a specific initial dataset; when every data point in the training dataset achieves an auto-similarity convergence exceeding a predefined threshold of $0.99$, the training is stopped and we can assume that the vector gradient's direction for each instance in the training dataset has stabilized.
 
\smallskip
The following step is to assess the similarity between cases within the training dataset and those outside it. This aims at identifying the most diverse cases among those not included in the training set, which will then be added to the training dataset to enhance diversity.  Firstly, the GNN runs for additional $10$ epochs to obtain the vector gradients of each out-of-training dataset instance. Then, a similarity matrix is computed by cross calculating the similarity between each in-training instance and each out-of-training instance. The case that shows the lowest similarity score is also the most diverse one and enables to promote diversification in the training dataset based on available data. Note that the values are normalized using a {z-score} value
\begin{equation}
z = \frac{S - \mu}{\sigma},
\end{equation}
where $S$ represents the similarity score for a specific data point, $\mu$ is the mean of all similarity scores, and $\sigma$ their standard deviation. In summary, the approach outlined here serves as a robust method for comparing and evaluating the similarities in data behaviour leveraging the neural network model, thereby enhancing the efficacy of the training process.
\end{appendices}

\section*{Acknowledgments}
The PhD fellowship of M.~Quattromini is supported by the Italian Ministry of University. This study has been partially funded under the National Recovery and Resilience Plan (NRRP), Mission 4 Component 2 Investment 1.3 - Call for tender No. 1561 of 11.10.2022 Project code PE0000021, Project title {Network 4 Energy Sustainable Transition – NEST}, funded by the European Union – NextGenerationEU. The support from the {Agence Nationale de la Recherche} (ANR) through grant ANR-21-REASON is gratefully acknowledged.

%Bibliography
\bibliographystyle{plainnat} % Stile della bibliografia

\bibliography{references}

\end{document}